\documentclass[sigconf]{acmart}
\AtBeginDocument{%
  \providecommand\BibTeX{{%
    \normalfont B\kern-0.5em{\scshape i\kern-0.25em b}\kern-0.8em\TeX}}}

\usepackage{multirow}
\usepackage{subcaption}
\usepackage{amsfonts,mathtools}

\usepackage{fixmath}
\newcommand{\overbar}[1]{\mkern 1.5mu\overline{\mkern-1.5mu#1\mkern-1.5mu}\mkern 1.5mu}
\copyrightyear{2022}
\acmYear{2022}
\setcopyright{acmcopyright}\acmConference[WWW '22]{Proceedings of the ACM Web Conference 2022}{April 25--29, 2022}{Virtual Event, Lyon, France}
\acmBooktitle{Proceedings of the ACM Web Conference 2022 (WWW '22), April 25--29, 2022, Virtual Event, Lyon, France}
\acmPrice{15.00}
\acmDOI{10.1145/3485447.3511972}
\acmISBN{978-1-4503-9096-5/22/04}

\begin{document}
\fancyhead{}
%%
%% The "title" command has an optional parameter,
%% allowing the author to define a "short title" to be used in page headers.
\title{StruBERT: Structure-aware BERT for 
%Keyword- and Content-Based 
%Data 
Table Search and Matching %and Similarity
}

%%
%% The "author" command and its associated commands are used to define
%% the authors and their affiliations.
%% Of note is the shared affiliation of the first two authors, and the
%% "authornote" and "authornotemark" commands
%% used to denote shared contribution to the research.
\author{Mohamed	Trabelsi}
\email{mot218@lehigh.edu}
\affiliation{%
 \institution{Lehigh University}
  %\streetaddress{113 Research Drive (Building C)}
  \city{Bethlehem}
  \state{PA}
  \country{USA}
  %\postcode{18015}
}

\author{Zhiyu Chen}
\email{zhc415@lehigh.edu}
\affiliation{%
  \institution{Lehigh University}
  %\streetaddress{113 Research Drive (Building C)}
  \city{Bethlehem}
  \state{PA}
  \country{USA}
  %\postcode{18015}
}

\author{Shuo Zhang}
\email{szhang611@bloomberg.net}
\affiliation{%
  \institution{Bloomberg}
  %\streetaddress{113 Research Drive (Building C)}
  %\city{Bethlehem}
  %\state{PA}
  \country{United Kingdom}
  %\postcode{18015}
}

\author{Brian D.\ Davison}
\email{davison@cse.lehigh.edu}
\affiliation{%
  \institution{Lehigh University}
  %\streetaddress{113 Research Drive (Building C)}
  \city{Bethlehem}
  \state{PA}
  \country{USA}
  %\postcode{18015}
}

\author{Jeff Heflin}
\email{heflin@cse.lehigh.edu}
\affiliation{%
  \institution{Lehigh University}
  %\streetaddress{113 Research Drive (Building C)}
  \city{Bethlehem}
  \state{PA}
  \country{USA}
  %\postcode{18015}
}

%%
%% By default, the full list of authors will be used in the page
%% headers. Often, this list is too long, and will overlap
%% other information printed in the page headers. This command allows
%% the author to define a more concise list
%% of authors' names for this purpose.
%\renewcommand{\shortauthors}{Trovato and Tobin, et al.}

%%
%% The abstract is a short summary of the work to be presented in the
%% article.
\begin{abstract}
A large amount of information is stored in data tables. Users can search for data tables using a keyword-based query. %representing an information need. 
A table is composed primarily of data values that are organized in %a 2D matrix with 
rows and columns providing implicit structural information. A table is usually accompanied by secondary information such as the caption, page title, etc., that form the textual information. Understanding the connection between the textual and structural information is an important yet neglected aspect in table retrieval as previous methods treat each source of information independently. %and focus more on the secondary textual information. 
In addition, users can search for data tables that are similar to an existing table, and this setting can be seen as a content-based table retrieval. %where the query and queried object are both data tables. 
In this paper, we propose StruBERT, a structure-aware BERT model that fuses the textual and structural information of a data table to produce context-aware representations for both textual and tabular content of a data table. StruBERT features are integrated in a new end-to-end neural ranking model to solve three table-related downstream tasks: keyword- and content-based table retrieval, and table similarity. We evaluate our approach using three datasets, and we demonstrate substantial improvements in terms of retrieval and classification metrics over state-of-the-art methods.
\end{abstract}

%%
%% The code below is generated by the tool at http://dl.acm.org/ccs.cfm.
%% Please copy and paste the code instead of the example below.
%%
\begin{CCSXML}
<ccs2012>
 <concept>
  <concept_id>10010520.10010553.10010562</concept_id>
  <concept_desc>Computer systems organization~Embedded systems</concept_desc>
  <concept_significance>500</concept_significance>
 </concept>
 <concept>
  <concept_id>10010520.10010575.10010755</concept_id>
  <concept_desc>Computer systems organization~Redundancy</concept_desc>
  <concept_significance>300</concept_significance>
 </concept>
 <concept>
  <concept_id>10010520.10010553.10010554</concept_id>
  <concept_desc>Computer systems organization~Robotics</concept_desc>
  <concept_significance>100</concept_significance>
 </concept>
 <concept>
  <concept_id>10003033.10003083.10003095</concept_id>
  <concept_desc>Networks~Network reliability</concept_desc>
  <concept_significance>100</concept_significance>
 </concept>
</ccs2012>
\end{CCSXML}
\ccsdesc[500]{Information systems~Retrieval models and ranking}
\ccsdesc[500]{Information systems~Structured text search}

\keywords
%{Table retrieval, table search, table similarity, pretrained language model, neural networks, learning to rank, information retrieval}
{table matching, table search, table similarity}

\maketitle

\section{Introduction}

\iffalse
Many datasets are publicly available for users to explore information across a variety of fields. Among all types of publicly available datasets, data tables represent the most prevalent form of data.
\fi
%A data table has multiple rows and columns. %These data tables contain vast amounts of information 
 %that are related to scientific, political, and cultural topics. Many users have questions that can be resolved from this data, but these questions may go unanswered due to ignorance
%regarding the presence of the data, ignorance regarding where to look for the data, and inability to formulate queries using the domain-specific vocabulary of the data's owners. 
Researchers have focused on utilizing the knowledge contained in tables in multiple tasks including augmenting tables \cite{Bhagavatula:2013:MEM:2501511.2501516, Zhang:2017:ESA,Zhang:2019:ADC,chen2018generating,yi2018recognizing}, 
extracting knowledge from tables \cite{Munoz:2014:ULD:2556195.2556266}, table retrieval  \cite{Cafarella:2009:DIR:1687627.1687750, Cafarella:2008:WEP:1453856.1453916,Pimplikar:2012:ATQ:2336664.2336665, nguyen2015,shraga,chen2020leveraging,chen2021wtr}, 
and table type classification ~\cite{Crestan2010AFT,crestan2011}.
\iffalse
\begin{figure*}[t!]
    \centering
    \begin{subfigure}[t]{0.5\textwidth}
        \centering
        \includegraphics[height=1.5in,width=3.9in]{table_matching_cases2.pdf}
        %\vspace*{-3mm}
        \caption{Table matching for content-based table retrieval and table similarity}
    \end{subfigure}%
    \begin{subfigure}[t]{0.45\textwidth}
        \centering
        \includegraphics[height=1.5in]{query_by_keywords.pdf}
        %\vspace*{-3mm}
        \caption{Keyword-based table retrieval}
    \end{subfigure}
    \vspace*{-2mm}
    \caption{Multiple scenarios for table matching and keyword-based table retrieval. Row- or/and column-based matching capture the semantic similarity between tables. Keyword-based queries can match table rows as in the row-based query example, or columns as in the column-based query example.}
    \label{scenarios}
\end{figure*}
\fi
\begin{figure*}[!t]
\centerline {\includegraphics[width=5.5in]{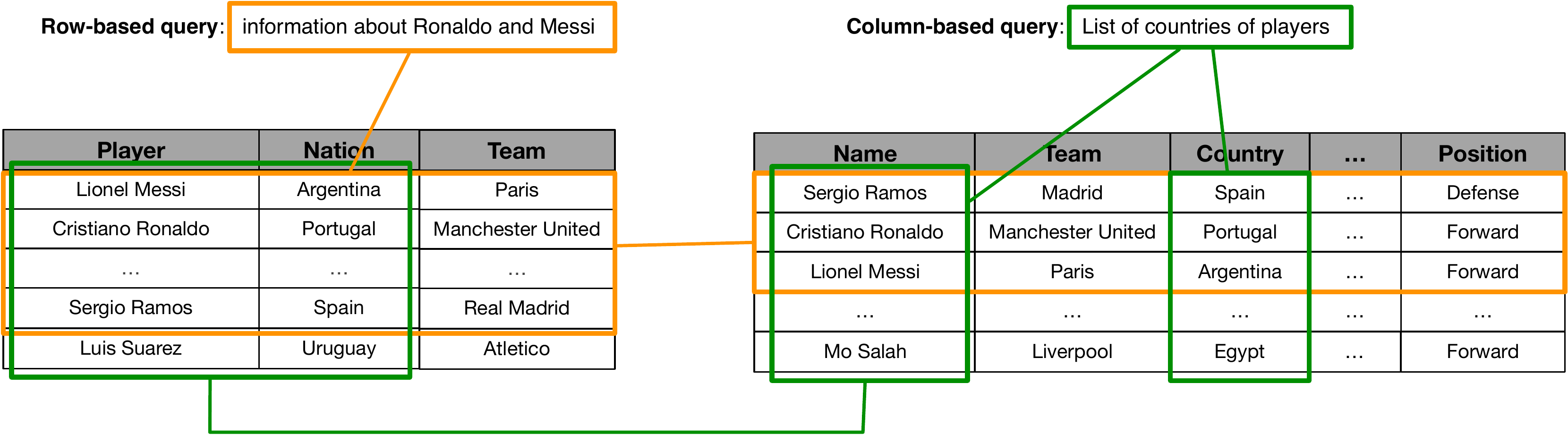}}
\vspace{-0.25in}
\vskip +0.1 true in \caption{\iffalse Multiple scenarios for table matching and keyword-based table retrieval.\fi Row-based matching (in orange) and column-based matching (in green) can happen between a keyword-based query and a table or between two tables.
%Keyword-based queries can match table rows  and/or columns. Row- and/or column-based matching captures the semantic similarity between tables. 
}
\label{scenarios}
\end{figure*}
Users can search for data tables using a keyword-based query as in document retrieval. \iffalse This scenario can be seen as ad-hoc table retrieval where multiple methods have been proposed in the literature. \fi  Additionally, users can look for similar data tables on the web to an existing table. This can be seen as a query by example scenario or content-based table retrieval. %, which is similar to content based image retrieval (CBIR) \cite{dharani2013,Zhou2017RecentAI}, where the query and the object queried are both data tables (instead of mages in CBIR). 
Content-based table retrieval requires a table matching phase to predict the semantic similarity between the query table and queried table. Another table-related task, that requires a table matching phase, is table similarity \cite{tabsim} in which the objective is to predict a binary semantic similarity between two tables. A table similarity algorithm can be used as a core component in multiple tasks %\cite{table_classification,Yoshida01amethod,table_fusion,DasSarma:2012:FRT:2213836.2213962}. 
 such as table classification and clustering \cite{table_classification,Yoshida01amethod}, table fusion \cite{table_fusion} and finding related tables \cite{DasSarma:2012:FRT:2213836.2213962}.  
%Google Fusion Tables \cite{DasSarma:2012:FRT:2213836.2213962} is an example of systems that computes the similarity between tables to find related tables in a large corpus of heterogeneous data. 
We consider content-based table retrieval and table similarity as two instances of table matching.
Prior methods \cite{DBLP:journals/corr/abs-1802-06159,Chen2020TableSU} ignore the structural information as data values are linearized, and treated as a single text. %The row and column vectors, that are used to capture the structural information by Trabelsi et al. \cite{dsrmm}, are context-independent and unable to capture the relationship between the structured form of a data table, and the textual form of both metadata and queries. 
 In table matching, Habibi et al.~\cite{tabsim} decouples the textual information from the structural information of a data table.%, and this can lead to context-free representations for both the textual and tabular content. 
 \iffalse Given the absence of inter-field interactions %when extracting the individual representations for the textual and tabular content, 
in feature extraction, there is a risk of losing important matching signals in the matching layer, and overfitting to a single field. \fi

 A table is composed of structural information defined by rows and columns, so that we  estimate the table matching score based on the semantic matching between rows and columns from both tables. Row-based matching selects mainly candidate tables that can be considered as additional records to the query table. On the other hand, column-based matching identifies %mainly 
tables that can potentially be used for table augmentation by join operations. In both cases, we have a structured query that is matched against structured tables. Users can also search for tables that match a sequence of %several 
keywords composed of several values, attributes, metadata, etc. Figure~\ref{scenarios} depicts both row/column-based matching between tables and row/column-based queries. In the former case, Figure \ref{scenarios} shows how columns and rows are matched to capture the semantic similarity between tables. In the latter case, Figure \ref{scenarios} shows two examples of keyword-based table retrieval where the query is a simple and unstructured natural language sequence. The row-based query is relevant to multiple rows of a table, and the column-based query contains keywords related to a subset of attributes from a table. 
%The query matches mainly the headers of the candidate table.

In order to overcome the limitations of prior methods in table retrieval and similarity, we propose a new model, called \textit{\textbf{Stru}cture-aware \textbf{BERT}} (StruBERT), that fuses the textual and structural information of a data table to produce a context-aware representation for both textual and tabular content of a table. In general, a table can be viewed as a row- and column-based structure, and rows and columns should contribute to both (1) the relevance score in table matching where rows and columns of a table pair are matched, and (2) the retrieval score in keyword-based table retrieval where table content is considered as a relevant field to the keywords of a query. Based on the observations from matching cases in Figure \ref{scenarios}, we propose a unified model that produces both row- and column-based features to predict the semantic matching between structured/unstructured query and structured 
\iffalse
candidate
\fi
table.  Inspired by TaBERT \cite{tabert}, which computes a joint representation for table columns and utterance tokens using vertical self-attention, % for the semantic parsing task,  
 we propose a horizontal self-attention that produces a joint representation for table rows and query tokens. %In other words, 
  Our proposed model produces four feature vectors that correspond to the joint representations of the structural and textual information of a table. Two fine-grained features represent the context-aware embeddings of rows and columns, where both horizontal and vertical attentions are applied over the column- and row-based sequences, respectively. 
\iffalse
The formed  sequences incorporate both the structural and textual information, and are encoded using BERT.
\fi
Two coarse-grained features capture the textual information from both row- and column-based views of a data table. These features are incorporated into a new end-to-end ranking model, called miniBERT, that is formed of one layer of Transformer \cite{transformer} blocks, and operates directly on the embedding-level sequences formed from StruBERT features to capture the cross-matching signals of rows and columns. 
\iffalse
in order to predict the relevance score.
\fi
%Using StruBERT features and the miniBERT ranking model, we solve three downstream tasks which are keyword- and content-based table retrieval, and table similarity.

In summary, we make the following contributions: (1) We propose a new structure-aware BERT model, called StruBERT, that introduces the idea of horizontal self-attention and fuses the structural and textual information of a data table to produce four context-aware features: two fine-grained structure- and context-aware representations for rows and columns, and two coarse-grained representations for row- and column-guided [CLS] embedding. (2) We propose a new ranking model, called miniBERT, that operates directly on the embedding-level sequences formed from StruBERT features to solve three table-related downstream tasks: keyword- and content-based table retrieval, and table similarity. (3) We evaluate over three datasets, and demonstrate that our new method outperforms the state-of-the-art baselines, and generalizes to multiple table-related downstream tasks.

\section{Related Work}

For supervised ranking of tables, multiple query, table, and query-table features are proposed in the literature \cite{Cafarella:2008:WEP:1453856.1453916,Bhagavatula:2013:MEM:2501511.2501516}. Zhang and Balog \cite{DBLP:journals/corr/abs-1802-06159} proposed extending these features with semantic matching between queries and tables using semantic spaces.
\iffalse
including: Word embeddings, Graph embeddings, Bag-of-entities and Bag-of-categories. 
 DBpedia \cite{Lehmann2015DBpediaA} %knowledge base 
is used to construct a vector of zeros and ones for both bag-of-entities and bag-of-categories. The dimension of bag-of-entities is equal to the total number of entities in the knowledge base, where a value of 1  indicates that % there is a link between entities in the knowledge base.
the entity is mentioned in the table.
The same applies to bag-of-categories with a dimension that is equal to the total number of Wikipedia categories.
\fi
\iffalse
A supervised model is then trained using the semantic and traditional features.
\fi
Recent works have used embedding techniques to learn a low dimensional representation for table tokens. %in different Natural Language Processing (NLP) tasks. 
%in table retrieval. 
Deng et al. \cite{table2vec} proposed a natural language modeling-based approach to create embeddings for table tokens. %embed table tokens into low dimensional vectors. 
The trained embedding is then used with entity similarity from a knowledge base to rank tables. Trabelsi et al.~\cite{Trabelsi} proposed a new word embedding model for the tokens
of table attributes %, called MCON, 
using the contextual information of every
table. 
\iffalse
Different formulations for contexts used to create the embeddings are proposed. 
%The authors argued that the different types of context should not all be treated uniformly and showed that data values are useful in creating a meaningful semantic representation of the attribute. 
%In addition to word embeddings, 
The model is used to predict additional contexts of each table that are incorporated into a mixed ranking model to compute %query-table 
the relevance score.
\fi
%The authors proposed using a mixed ranking model that incorporates the metadata of a table and additional predicted contexts in order to predict query-table relevance score in the table retrieval task. %Using multiple and differentiated contexts leads to more useful attribute embeddings for the table retrieval task. 
%Using matrix factorization, Chen et al. \cite{chen2019smlr} generated additional headers that are used in ranking table-query pairs. The authors showed that the generated headers improve the performance of unsupervised table retrieval.

Deep contextualized language models, like BERT \cite{Devlin2019BERTPO} and RoBERTa \cite{Liu2019RoBERTaAR}, have been recently proposed 
to solve multiple tasks \cite{wang-etal-2018-glue,liu-etal-2019-multi,Yilmaz2019ApplyingBT,dai2019,Nogueira2019MultiStageDR,Yang2019SimpleAO,sakata2019,Nogueira2019PassageRW,selab_arxiv,survey_doc_retrieval,selab_ijcnn}. 
% natural language understanding \cite{wang-etal-2018-glue,liu-etal-2019-multi} and information retrieval \cite{Yilmaz2019ApplyingBT,dai2019,Nogueira2019MultiStageDR,Yang2019SimpleAO,sakata2019,Nogueira2019PassageRW} tasks. %Different from traditional word embeddings, the pre-trained neural language models are contextual with the representation of a token is a function of the entire sentence. This is mainly achieved using a self-attention structure called transformer \cite{transformer}.
% \iffalse
% to solve multiple tasks in natural language understanding \cite{wang-etal-2018-glue,liu-etal-2019-multi} and information retrieval including document retrieval \cite{dai2019,Nogueira2019MultiStageDR,Yang2019SimpleAO}, frequently asked question retrieval \cite{sakata2019}, passage re-ranking \cite{Nogueira2019PassageRW}, and table retrieval \cite{Chen2020TableSU}. 
% \fi
Building on BERT, Chen et al.~\cite{Chen2020TableSU} proposed a BERT-based ranking model to capture the matching signals between the query and the table fields using the sentence pair setting. They first select the most salient items of a table to construct the BERT representation, where different types of table items and salient signals are tested. 
\iffalse
Neural ranking models have been leveraged in document retrieval, and have led to significant improvement in document ranking and retrieval results.
\fi
Trabelsi et al.~\cite{dsrmm} have shown that neural ranking models can be used in table retrieval by proposing a deep semantic and relevance matching model (DSRMM).  %This ranking model captures semantic and relevance signals between the query and table. 
%The structural information of a data table is incorporated into DSRMM by computing row and column summary vectors.
Shraga et al.~\cite{shraga2020web} use neural networks to learn unimodal features of a table which are combined into a multimodal representation. %The final table-query relevance is estimated based on the query representation and multimodal representation.  
Tables can also be represented as graphs to solve table retrieval \cite{gtr,multiemrgcn,chen2021mgnets}.

Table similarity consists of predicting the semantic similarity between tables and then classifying a table pair as similar or dissimilar. %Content based table retrieval is related to table similarity where the relevance scores for table pairs are used to rank tables. 
Das Sarma et al.~\cite{DasSarma:2012:FRT:2213836.2213962} proposed a table similarity method that is based on entity consistency and expansion, and schema similarity, and is used to find related tables in a large corpus of heterogeneous data. 
\iffalse
In entity consistency, similar tables should have similar entities, and in entity expansion, the queried table should add new entities to the query table. In schema similarity, similar tables should have similar schemas that represent similar entities.
Relevance scores are computed for entity consistency and expansion, and schema similarity, and then combined to predict the entity complement score of a table pair.
\fi
 Deep learning models have been leveraged to predict the similarity score between tables. %TabSim \cite{tabsim} is composed of two separate deep learning models that extract the textual and structural features of a table. Then, a Euclidean-based Siamese model is used to predict the semantic relatedness of a table pair. 
TabSim \cite{tabsim} treats the data table fields independently in which one bi-LSTM model is used to map the caption of a data table to an embedding vector, and a second attention-based model is used to compute the embeddings of the columns of a data table. %The caption and column embeddings are concatenated to form the representation of a data table. 
%After extracting the feature vectors of two data tables, 
%Euclidean distance is used to predict the semantic similarity between tables.

\begin{figure*}[!t]
\centerline {\includegraphics[width=5.3in]{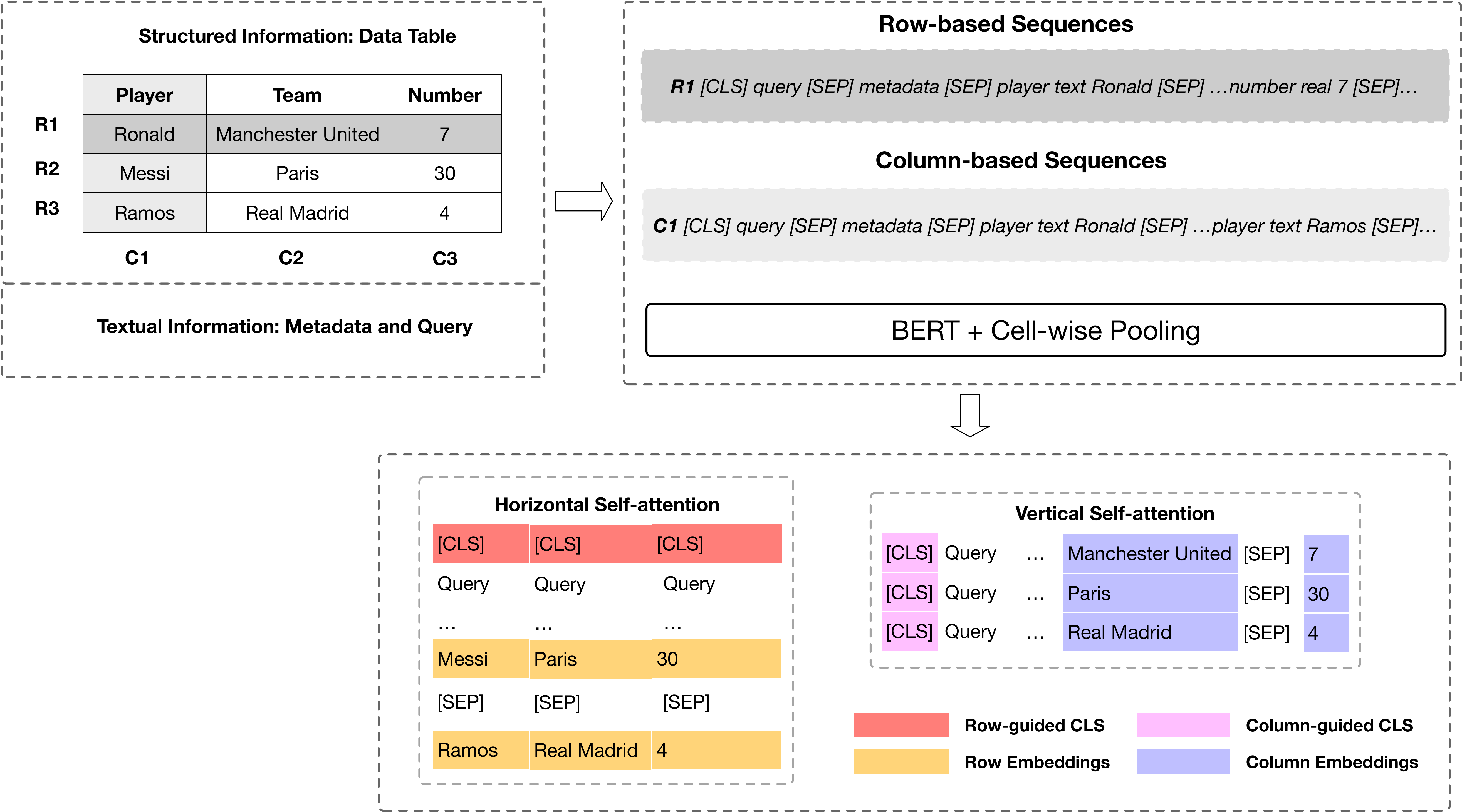}}
% {StruBERT_review_simplified.pdf}}
\vspace{-.2in}
\vskip +0.1 true in \caption{%Architecture of StruBERT. 
%The bottom box presents the steps of forming the input sequences to StruBERT. The upper box presents the architecture of StruBERT.
Column- and row-based sequences are formed from the structural and textual information of the table. The sequences are encoded using BERT+cell-wise pooling. Horizontal and vertical self-attentions are applied to the encoded column- and row-based sequences, respectively to obtain four feature vectors: two fine-grained features (row and column embeddings), and two coarse-grained features (row- and column-guided [CLS] embeddings). 
}
\label{strubert}
\end{figure*}

\section{Problem Statement}

We formally define the three table-related downstream tasks that we address in this paper.

\subsection{Keyword-based Table Retrieval}

Given a keyword-based query $q=q_1 q_2 \ldots q_m$ where $m$ is  the  length  of the query and $q_i$ is the $i$-th  token  of $q$, the objective is to find a relevant set of tables from a table corpus $\mathcal{C}=\{T_1, T_2,\ldots,T_l\}$, with $l$ is the total number of data tables. Our model takes as input a query-table pair $(q,T_j), j=1,2,\ldots,l$, and produces a real-valued relevance score for each pair, where these scores are used to rank data tables against the user's query.

\subsection{Content-based Table Retrieval}

In content-based table retrieval, the user searches for similar data tables on the web to an existing table. So, the query $q$ is also a data table $T_i$ ($q=T_i$). In this setting, our model takes as input a query-table pair $(T_i,T_j), j=1,2,\ldots,l$, and produces a real-valued relevance score for each pair, where these scores are used to rank data tables against the table-based user's query. %With binary relevance score, content-based table retrieval can be seen as table similarity.

\subsection{Table Similarity}

Similar to content-based table retrieval, our model takes as input pairs of tables. However, our model outputs a binary label, instead of a real-valued relevance score, for each table pair in order to classify table pairs as similar or dissimilar.

\textbf{We consider content-based table retrieval and table similarity as two instances of table matching} because tables should be matched either to compute the retrieval score in content-based table retrieval, or the classification score in table similarity.

\section{StruBERT: Representation Learning}

In this section, we introduce our proposed method StruBERT that fuses structural and textual information of a data table to produce structure- and context-aware features. These features are used in downstream tasks that are related to data table search and matching. 
% We first describe how to form the input sequences to StruBERT from the structural and textual information of a data table. Then, we describe how StruBERT produces multiple context-aware features that incorporate both the structural and textual information. Finally, we show how to use these features in an end-to-end ranking model for table-related downstream tasks.

\subsection{Table Views}
%The bottom box of Figure \ref{strubert} presents the steps of forming the input sequences to StruBERT. 
The primary input to our model is a table $T_j$ that has $s$ rows and $l$ columns. Each table has two forms of information. The first form is the structural information %, denoted by $S_j$, 
which is composed of headers and data values. A table can be seen as a 2D matrix of cells, and for the purpose of explanation, we assume that the first row corresponds to the headers $c_1,c_2,\ldots,c_l$, and the remaining $s-1$ rows are data values. The $i$-th column of $T_j$ has the values $v_{1i},v_{2i},\ldots,v_{(s-1)i}$. The second form of information is the textual information which corresponds to the context fields of a table. Several text fields can be used to describe the table such as the caption, the title of the page and section that contain the table, etc. We denote these context fields by the metadata which forms the textual information of a table. In the case of keyword-based table retrieval, the query is considered as an additional form of textual information because the final representations of StruBERT should capture early interactions between the table and query as in interaction-based retrieval models \cite{Hu2014ConvolutionalNN,Guo2016ADR,DeepRank} that have achieved better results than the representation-based models \cite{Nie2018EmpiricalSO}. By learning early interactions between the table and keyword-based query, StruBERT produces structure- and context-aware features, where the query is part of the context. %We describe the main components of StruBERT.

\iffalse A data table can be seen as a column- and row-based structure. \fi As shown in Figure \ref{strubert}, we propose forming two sets of sequences, denoted by column- and row-based sequences, that are formed on the basis of column- and row-based views, respectively, of a given data table. \citet{tabert} proposed a row linearization to form sequences from a data table in order to solve the semantic parsing over tables task. Inspired by that, we incorporate row linearization to form row-based sequences, and we propose a column linearization to form column-based sequences. 

Given that $T_j$ has $l$ columns, we form $l$ column-based sequences. The $i$-th column-based sequence is given by:
\begin{equation}
\tilde{c_i}=c_it_iv_{1i}[SEP]c_it_iv_{2i}[SEP]\ldots [SEP]c_it_iv_{(s-1)i}[SEP]
\end{equation}
where $t_i \in [real,text]$ is the type of $c_i$. For example, the first column in the data table shown in Figure \ref{strubert} has a type $text$, and the third column has a type $real$. We use the first column of the table in Figure \ref{strubert} to illustrate an example of a column-based sequence:
$$player \textrm{ } text \textrm{ } Ronaldo \textrm{ } [SEP]\textrm{ }  player \textrm{ } text \textrm{ } Messi \textrm{ } [SEP] \textrm{ } \ldots$$
We denote the set of column-based sequences by $\tilde{\mathcal{C}}=\{\tilde{c_1},\tilde{c_2},\ldots,\tilde{c_l}\}$.
Similarly, we form $s-1$ row-based sequences. The $i$-th row-based sequence is given by: 
\begin{equation}
\tilde{r_i}=c_1t_1v_{i1}[SEP]c_2t_2v_{i2}[SEP]\ldots [SEP]c_lt_lv_{il}[SEP]
\end{equation}
We use the first row of the data table in Figure \ref{strubert} to illustrate an example of a row-based sequence:
$$player \textrm{ } text \textrm{ } Ronaldo \textrm{ } [SEP]\textrm{ }  team \textrm{ } text \textrm{ } Manchester \textrm{ } United \textrm{ } [SEP] \textrm{ } \ldots$$
We denote the set of row-based sequences by $\tilde{\mathcal{R}}=\{\tilde{r_1},\tilde{r_2},\ldots,\tilde{r_{(s-1)}}\}$. $\tilde{\mathcal{C}}$ and $\tilde{\mathcal{R}}$ capture only the structural information of $T_j$. To incorporate the textual information into the structure-based sequences, we concatenate the textual information with each sequence from the structure-based sequences $\tilde{\mathcal{C}} \cup \tilde{\mathcal{R}}$ using the [CLS] and [SEP] tokens of BERT. Given that the textual information $Te_j$ of $T_j$ is formed of $p$ fields $f_1,f_2,\ldots,f_p$, the new structure- and context-aware sequences are given by:
\begin{equation}
\overbar{c_i}=[CLS]\widetilde{Te_j}[SEP]\tilde{c_i}[SEP]
\end{equation}
\begin{equation}
\overbar{r_i}=[CLS]\widetilde{Te_j}[SEP]\tilde{r_i}[SEP]
\end{equation}
where:
\begin{equation}
\widetilde{Te_j}=f_1[SEP]f_2[SEP]\ldots[SEP]f_p
\end{equation}
We denote the column- and row-based structure-aware sequences by $\overbar{\mathcal{C}}=\{\overbar{c_1},\overbar{c_2},\ldots,\overbar{c_l}\}$, and  $\overbar{\mathcal{R}}=\{\overbar{r_1},\overbar{r_2},\ldots,\overbar{r_{(s-1)}}\}$, respectively.

\subsection{StruBERT Model}

Figure \ref{strubert} presents  StruBERT which is composed of two phases: sequence encoding and self-attention over encoded sequences.
\subsubsection{Sequence Encoding}

To capture the dependencies between the textual information and data values in each sequence from $\overbar{\mathcal{C}} \cup \overbar{\mathcal{R}}$, BERT is used as a sequence encoder which produces contextualized embeddings for each token in the tokenized sequence using BERT tokenizer. BERT is preferred over a recurrent architecture because BERT is composed of Transformer blocks that capture long-range dependencies with self-attention better than  recurrent architectures~\cite{transformer}, and is pretrained on large textual data.

After row (column) linearization and BERT tokenization, each cell has multiple tokens. To compute a single embedding for each cell, we incorporate cell-wise average pooling \cite{tabert} after the BERT encoding step to pool over the contextualized tokens for each cell defined by [header\_name type cell\_content]. BERT is composed of $L$ layers of Transformer blocks. The cell-wise average pooling is applied on the contextualized embedding that is obtained from the last layer. The contextualized embedding of the column-based sequence $\overbar{c_i}$ is given by:
\begin{equation}
\footnotesize \overbar{\mathbold{c_i}}=\mathbold{[CLS]\widetilde{Te_j}[SEP]v_{1i}[SEP]\ldots [SEP]v_{(s-1)i}[SEP]}
\end{equation}
where:
\begin{equation}
 \mathbold{v_{ki}}= \frac{\sum_{w \in BertTok(c_it_iv_{ki})} \mathbold{h_w^L} }{|BertTok(c_it_iv_{ki})|};\hspace{0.2cm} k=1,2,\ldots,s-1
\end{equation}
$BertTok(c_it_iv_{ki})$ represents the tokens that are obtained after tokenizing the sequence $c_it_iv_{ki}$ using BERT tokenizer, and $\mathbold{h_w^L} \in \mathbb{R}^{d}$ is the contextualized embedding of dimension $d$ from the $L$-th layer of BERT for the token $w \in BertTok(c_it_iv_{ki})$. Similarly, the cell-wise average pooling is used to compute the contextualized embedding for the row-based sequence $\overbar{r_i}$, denoted by $\overbar{\mathbold{r_i}}$.
\iffalse
which is given by:
\begin{equation}
\footnotesize \overbar{\mathbold{r_i}}=\mathbold{[CLS]Metadata[SEP]v_{i1}[SEP]\ldots [SEP]v_{il}[SEP]}
\end{equation}
where:
\begin{equation}
 \mathbold{v_{ik}}= \frac{\sum_{w \in BertTokenizer(c_kt_kv_{ik})} \mathbold{h_w^L} }{|BertTokenizer(c_kt_kv_{ik})|} ;\hspace{0.2cm} k=1,2,\ldots,l
\end{equation}
\fi

We denote the column- and row-based contextualized embeddings that are obtained after BERT and cell-wise average pooling by $\boldsymbol{\overbar{\mathcal{C}}}=\{\mathbold{\overbar{c_1}},\mathbold{\overbar{c_2}},\ldots,\mathbold{\overbar{c_l}}\}$ and $\boldsymbol{\overbar{\mathcal{R}}}=\{\mathbold{\overbar{r_1}},\mathbold{\overbar{r_2}},\ldots,\mathbold{\overbar{r_{(s-1)}}}\}$, respectively.

\subsubsection{Horizontal and Vertical self-attention}
Self-attention is incorporated into StruBERT for two reasons. First, %the contextualized embeddings in $\boldsymbol{\overbar{\mathcal{C}}} \cup \boldsymbol{\overbar{\mathcal{R}}}$ capture independent column- and row-level structural and textual information, and ignore the row- and column-level dependency in the sequences from $\overbar{\mathcal{C}}$ and $\overbar{\mathcal{R}}$, respectively. %The main elements of a data table are rows and columns, so that the main outputs of our proposed model are row and column embeddings.
the contextualized embeddings in $\boldsymbol{\overbar{\mathcal{C}}}$ capture independent column-level structural and textual information, and ignore the row-level dependency as a result of tabular structure. The same conclusion applies for $\overbar{\mathcal{R}}$ where column-level dependency is not captured in the row-level embeddings.
Second, cell values are not equally important for the representations of rows and columns. We incorporate vertical self-attention \cite{tabert} to operate over row-based embeddings to produce column embeddings, and we propose a horizontal self-attention that operates over column-based embeddings to form row embeddings. Both attentions are similar to the Transformer \cite{transformer}, and the naming of horizontal and vertical attention comes from the orientation of input sequences to attention blocks. 

\textbf{Horizontal self-attention}: To capture the row-level dependency between the column-based contextualized embeddings of $\boldsymbol{\overbar{\mathcal{C}}}$, we propose a multi-head horizontal self-attention that operates on horizontally aligned tokens from the column-based embeddings as shown in Figure \ref{strubert}. The horizontal self-attention is formed of $H$ layers of Transformers, and we use the output of the last layer as the row-level self-attention representation. We produce two types of features from the horizontal self-attention step after applying row-level average pooling. First, we obtain $s-1$ row embeddings which can be seen as fine-grained structure- and context-aware features. Second, by averaging the [CLS] embedding from each column, we produce a row-guided [CLS] which represents a coarse-grained structure and context-aware feature. In conclusion, the horizontal self-attention features are based on interpreting the data table as a column-based structure, followed by row-level dependency.

\textbf{Vertical self-attention}: Similarly, a data table can be interpreted as a row-based structure, followed by column-level dependency. In this case, $V$ layers of vertical self-attention \cite{tabert} operate on the row-based contextualized embeddings of $\boldsymbol{\overbar{\mathcal{R}}}$. We also obtain two types of features from the vertical self-attention. First, we obtain $l$ fine-grained column embeddings by averaging the last output of the vertical self-attention over the vertically aligned tokens from the row-based embeddings. Second, we obtain a coarse-grained column-guided [CLS] embedding that interprets the data table as a row-based structure, followed by column-level dependency.

In conclusion, StruBERT generates four structure- and context-aware features: two fine-grained features which are the contextualized row and column embeddings, denoted by $\boldsymbol{E_r} \in \mathbb{R}^{(s-1) \times d}$ and $\boldsymbol{E_c} \in \mathbb{R}^{l \times d}$, respectively,  and two coarse-grained features which are the row- and column-guided [CLS] embeddings, denoted by $\boldsymbol{[CLS]_r} \in \mathbb{R}^{d}$ and $\boldsymbol{[CLS]_c} \in \mathbb{R}^{d}$, respectively.

\section{StruBERT in Downstream Tasks}
\iffalse
We integrate StruBERT into end-to-end architectures to solve table-related downstream tasks. We denote an end-to-end model by $g=N\circ F$, where $F$ is a StuBERT-based feature extractor and $N$ is a ranking model that maps the extracted features to classification or retrieval score depending on the task. In this section, we address the tasks of %keyword and content based data table search 
table search and table matching via end-to-end neural models where StruBERT is integrated as a feature extractor. %We keep the same name StruBERT for the end-to-end models as the proposed ranking models are built on top of StruBERT features.
\fi

We integrate StruBERT as a feature extractor $F$ into end-to-end architectures to solve table-related downstream tasks. In this section, we address the tasks of 
table search and table matching, and we show how to map StruBERT features to a classification or retrieval score depending on the task. 

\subsection{Table Matching} \label{table_matching}

In table matching tasks, both the query and the queried object are data tables. The neural ranking model should capture the semantic similarity between the structural and textual information of a table pair in order to predict the relevance score. %The classification case of data tables is known as table similarity. In both cases, the inputs are two tables, and the output is a relevance score (real-valued or binary). 
To this end, we propose a Siamese \cite{Lecun}-based model that predicts the relevance score of a table pair $(T_i,T_j)$. In table matching, the textual information of each table contains only the metadata because the keyword-based query is absent. %Figure \ref{table_sim} shows an overview of our proposed 
%We propose a Siamese model to predict relevance scores for table pairs. 
Structure- and context-aware features are extracted from each table using StruBERT:
    $$F(T_i,T_j)=(StruBERT(T_i),StruBERT(T_j))$$
    $$F(T_i,T_j)=((\boldsymbol{E_r^i},\boldsymbol{E_c^i},\boldsymbol{[CLS]_r^i},\boldsymbol{[CLS]_c^i}),(\boldsymbol{E_r^j},\boldsymbol{E_c^j},\boldsymbol{[CLS]_r^j},\boldsymbol{[CLS]_c^j}))$$
After extracting features from each table using StruBERT, we obtain coarse- and fine-grained features for each table. We propose a ranking model \iffalse $N$ \fi that captures the semantic similarities within the fine-grained features ($(\boldsymbol{E_r^i},\boldsymbol{E_c^i})$ and $(\boldsymbol{E_r^j},\boldsymbol{E_c^j})$), and coarse-grained features ($(\boldsymbol{[CLS]_r^i},\boldsymbol{[CLS]_c^i})$ and $(\boldsymbol{[CLS]_r^j},\boldsymbol{[CLS]_c^j})$).
%In each category of features, there are both row and column based representations. %Inspired by the word-entity ranking model \cite{Xiong2017WordEntityDR} that includes interactions between word-based feature vectors and entity-based feature vectors based on the inter- and intra-space matching signals, we propose a cross-matching layer within each category of feature, where intra-space matching signals correspond to row-row and column-column matching, and inter-space matching signals correspond to row-column and column-row matching. By incorporating both the inter and intra-space matching signals, we predict the semantic similarity between data tables regardless of the table orientation inconsistency that can result from automatic table extraction.

\subsubsection{Cross-matching of Fine-grained Features:}
\begin{figure}[!t]
\centerline {\includegraphics[width=.5\textwidth]{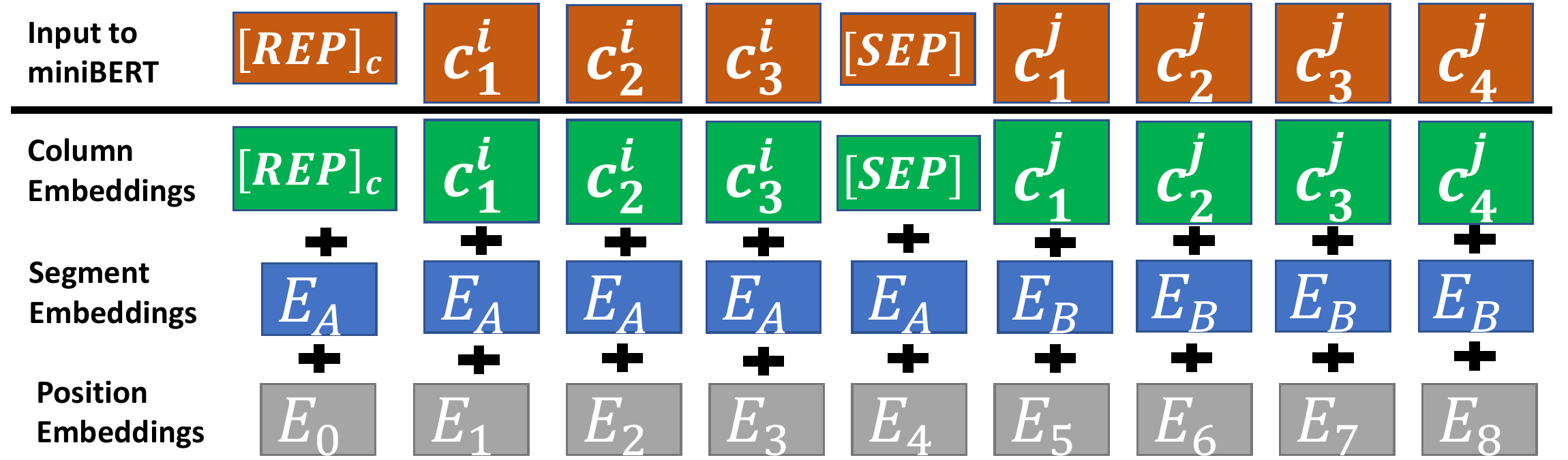}}
\vspace{-.2in}
\vskip +0.1 true in \caption{Embedding-level sequence input of miniBERT for cross-matching of columns. The input to miniBERT is the sum of %three embeddings:
 column, segment, and position embeddings. In this example, $\boldsymbol{E_c^i} \in \mathbb{R}^{3 \times d}$ is composed of $c^i_k \in \mathbb{R}^d,k \in [1,2,3]$, and $\boldsymbol{E_c^j} \in \mathbb{R}^{4 \times d}$ is composed of $c^j_k \in \mathbb{R}^d,k \in [1,2,3,4]$.
}
\label{input_miniBERT}
\end{figure}

To capture cross-matching signals of row and column embeddings for table pairs, we propose a model, called  miniBERT, that operates directly on the embedding-level sequences of fine-grained features of StruBERT. miniBERT is composed of three trainable vectors $\boldsymbol{[REP]_c} \in \mathbb{R}^{d}$, $\boldsymbol{[REP]_r} \in \mathbb{R}^{d}$, and $\boldsymbol{[SEP]} \in \mathbb{R}^{d}$, and 1 layer of Transformer blocks with 4 attention heads.\footnote{We tried to increase the number of layers and attention heads, but we did not notice an improvement in the reported evaluation metrics.} The input to miniBERT for the column-based cross-matching of a table pair $(T_i,T_j)$ is shown in Figure \ref{input_miniBERT}. $\boldsymbol{[REP]_c}$ is introduced to aggregate the matching signals between $\boldsymbol{E_c^i}$ and $\boldsymbol{E_c^j}$. We form the embedding-level sequence for the column embeddings of a table pair $(T_i,T_j)$:
\begin{equation}
M_{c_i c_j}=\boldsymbol{[REP]_c} \oplus \boldsymbol{E_c^i} \oplus \boldsymbol{[SEP]} \oplus \boldsymbol{E_c^j}
\end{equation}
where $\boldsymbol{[SEP]}$ is used to separate $\boldsymbol{E_c^i}$ and $\boldsymbol{E_c^j}$. As in BERT, we sum three different embeddings to obtain the input embeddings to miniBERT. As shown in Figure \ref{input_miniBERT}, in addition to the column embeddings, the segment embeddings are used to indicate the column embeddings that belong to $T_i$ and $T_j$, and the position embeddings are used to encode the position of each vector in $M_{c_i c_j}$. The position embedding of $\boldsymbol{[REP]_c}$ is in particular useful to indicate that the final hidden state from the first position aggregates the embedding-level sequence $M_{c_i c_j}$. So, miniBERT takes the embedding-level sequence, that is formed by summing the column, segment and position embeddings, as input, then miniBERT outputs the hidden state of $\boldsymbol{[REP]_c}$ from the Transformer block, denoted by $miniBERT(\boldsymbol{[REP]_c})$, that captures the  bidirectional cross-attention between $\boldsymbol{E_c^i}$ and $\boldsymbol{E_c^j}$.

Similarly, we use miniBERT to compute the hidden state of $\boldsymbol{[REP]_r}$, denoted by $miniBERT(\boldsymbol{[REP]_r})$, from the embedding-level sequence input for rows defined by:
\begin{equation}
M_{r_i r_j}=\boldsymbol{[REP]_r} \oplus \boldsymbol{E_r^i} \oplus \boldsymbol{[SEP]} \oplus \boldsymbol{E_r^j}
\end{equation}
There are mainly two advantages from using miniBERT as a ranking model on top of the StruBERT features. First, a row- or column-based permutation for a table does not change the meaning of the table. The self-attention of the Transformer blocks in miniBERT is particularly useful where each embedding attends to all embeddings in the column- and row-based embedding-level sequences regardless of the position information. %This makes miniBERT a permutation-invariant ranking model. 
 Second, evaluating the semantic similarity between tables is not based only on one-to-one mapping between columns or rows. For example, one column from $T_i$ can summarize the information that is present in three columns from $T_j$. The attention weights in the attention heads of miniBERT are valuable to capture many-to-many relationships between columns (rows) of a table pair by aggregating information both within and across table columns (rows).

\iffalse
\begin{align}
\begin{multlined}
 M_{r_i r_j}= \big(\boldsymbol{[REP]_r}+\boldsymbol{P_0}+\boldsymbol{S_A}\big)\oplus (\boldsymbol{E_r^i}+\boldsymbol{P_1^{|\boldsymbol{E_r^i}|}}+\boldsymbol{S_A^{|\boldsymbol{E_r^i}|}})\\+ (\boldsymbol{[SEP]}+\boldsymbol{P_{|\boldsymbol{E_r^i}|+1}}+\boldsymbol{S_A}) \oplus (\boldsymbol{E_r^j}+\boldsymbol{P_{|\boldsymbol{E_r^i}|+2}^{|\boldsymbol{E_r^i}|+1+|\boldsymbol{E_r^j}|}}+\boldsymbol{S_B^{|\boldsymbol{E_r^j}|}})
\end{multlined}
%\label{skipgram}
\end{align}

\fi

\subsubsection{Cross-matching of Coarse-grained Features:} Similarly to the fine-grained features, we construct the cross-matching features between the coarse-grained features of $T_i$ and $T_j$. We define the interaction vectors $\mathcal{F}=\{\boldsymbol{F_{r_ir_j}},\boldsymbol{F_{c_ic_j}}\}$, where $\boldsymbol{F_{r_ir_j}}$, and $\boldsymbol{F_{c_ic_j}}$ denote the interactions of $\boldsymbol{[CLS]_r^i}$-$\boldsymbol{[CLS]_r^j}$, and $\boldsymbol{[CLS]_c^i}$-$\boldsymbol{[CLS]_c^j}$, respectively, and the elements of each vector are computed using pointwise multiplication between the embeddings of the corresponding row- and column-guided [CLS]:
\begin{equation}
\begin{aligned}
\boldsymbol{F_{r_i r_j}} &=\boldsymbol{[CLS]_r^i} \odot \boldsymbol{[CLS]_r^j} \text{ ; } \boldsymbol{F_{c_i c_j}}=\boldsymbol{[CLS]_c^i} \odot \boldsymbol{[CLS]_c^j} 
\end{aligned}
\end{equation}
\iffalse
We forward the coarse-grained interaction vectors to a multi-layer perceptron (MLP) to reduce the dimensionality of the row- and column-guided [CLS]:
\begin{equation}
MLP(\mathcal{F})=MLP\left(F_{r_i r_j} \oplus F_{c_i c_j} \oplus F_{r_i c_j} \oplus F_{c_i r_j}\right)
\end{equation}
\fi

\subsubsection{Ranking Layer:} The fine- and coarse-grained features are used as input to a ranking layer to predict the relevance score of a table pair. 
\iffalse
$KP(\mathcal{M})$ and $MLP(\mathcal{F})$ have different range of values, so that we apply $\mathcal{L}_2$ normalization per interaction matrix to $KP(\mathcal{M})$. The final feature vector of a given table pair $(T_i,T_j)$ is given by:
\begin{equation}
\Phi (T_i,T_j)= \mathcal{L}_2^{norm}\left(KP(\mathcal{M})\right) \oplus MLP(\mathcal{F})
\end{equation}
\fi
The final feature vector of a table pair $(T_i,T_j)$ is given by:
\begin{equation}
\footnotesize
\Phi (T_i,T_j)=\boldsymbol{F_{r_i r_j}} \oplus \boldsymbol{F_{c_i c_j}} \oplus miniBERT(\boldsymbol{[REP]_r}) \oplus miniBERT(\boldsymbol{[REP]_c})
\end{equation}
A final linear layer is used to predict the relevance score of the table pair $(T_i,T_j)$ using $\Phi (T_i,T_j)$.
\iffalse
\begin{equation}
g(T_i, T_j)=\omega_{r}^{T} \Phi (T_i,T_j)+b_{r}
\end{equation}
where $\omega_{r}$ and $b_{r}$ are the parameters of the linear layer.
\fi

\subsection{Keyword-based Table Retrieval} \label{keyword_table_retrieval}

The query $q$ is composed of several keywords, $q_1 q_2 \ldots q_m$ %where $m$ is  the  length  of the query %and $q_l$ is the $l$-th  token  of $q$, 
 and the queried object is a data table $T_i$ from a table corpus $\mathcal{C}$. In addition to the table's metadata, the textual information $Te_i$ contains the query $q$ so that the outputs of StruBERT capture early interactions between the query, and the structural and textual information of a data table. We use the same notations of the table matching case, and we denote the outputs of StruBERT for a given query-table pair $(q,T_i)$ by: $\boldsymbol{E_r^i},\boldsymbol{E_c^i},\boldsymbol{[CLS]_r^i},\boldsymbol{[CLS]_c^i}$. 
We apply miniBERT to the single embedding-level sequences defined by: 
\begin{equation}
\begin{aligned}
M_{r_i q}=\boldsymbol{[REP]_r} \oplus \boldsymbol{E_r^i(q)} \oplus \boldsymbol{[SEP]} \\
M_{c_i q}=\boldsymbol{[REP]_c} \oplus \boldsymbol{E_c^i(q)} \oplus \boldsymbol{[SEP]}
\label{pool_feat}    
\end{aligned}
\end{equation}
where $\boldsymbol{E_r^i}$ and $\boldsymbol{E_c^i}$ are functions of $q$ because $q \in Te_i$ in the case of keyword-based table retrieval. We use the final hidden states of $\boldsymbol{[REP]_r}$ and $\boldsymbol{[REP]_c}$  %denoted by $miniBERT(\boldsymbol{[REP]_r})$ and $miniBERT(\boldsymbol{[REP]_c})$ respectively, 
that are obtained from miniBERT as the row- and column-based aggregate for the query-table pair $(q,T_i)$, respectively.
\iffalse
\textbf{Ranking layer:} 
\fi
A query-table pair $(q,T_i)$ is represented using four feature vectors: row and column outputs from miniBERT %($miniBERT(\boldsymbol{[REP]_r})$, $miniBERT(\boldsymbol{[REP]_c})$), 
and row- and column-guided [CLS] embeddings ($\boldsymbol{[CLS]_r^i},\boldsymbol{[CLS]_c^i}$). We concatenate these features to obtain the final representation for $(q,T_i)$, 
\iffalse
denoted by $\Phi (q,T_i)$.
\fi
which is used as input to a linear layer to predict the relevance score of the query-table pair $(q,T_i)$.
\iffalse
\begin{equation}
\footnotesize
\Phi (q,T_i)= miniBERT(\boldsymbol{[REP]_r}) \oplus miniBERT(\boldsymbol{[REP]_c}) \oplus \boldsymbol{[CLS]_r^i} \oplus \boldsymbol{[CLS]_c^i}
\end{equation}
\fi
\iffalse
As in table matching, a final linear layer is used to predict the relevance score of the query-table pair $(q,T_i)$ using $\Phi (q,T_i)$.
\fi
\iffalse
\begin{equation}
g(q, T_i)=\omega_{r}^{T} \Phi (q,T_i)+b_{r}
\end{equation}
where $\omega_{r}$ and $b_{r}$ are the parameters of the linear layer.
\fi
\section{Evaluation}

\subsection{Data Collections}
\subsubsection{WikiTables}

This dataset is composed of the WikiTables corpus \cite{Bhagavatula2015TabELEL} containing over $1.6M$ tables. Each table has five indexable fields: table caption, attributes, data rows, page title, and section title. 
\iffalse
In addition, each table contains statistics which are: number of columns, number of rows, and set of numerical columns of the table. Additional LTR features \cite{DBLP:journals/corr/abs-1802-06159} are provided for each table. The set of LTR features includes the number of in-links to the page embedding the table, number of out-links from the page embedding the table, number of page views, etc.
\fi
We use the same test queries that were used by Zhang
and Balog \cite{DBLP:journals/corr/abs-1802-06159}. %The queries are divided into two subsets: the first contains queries collected by Cafarella et al. \cite{Cafarella:2009:DIR:1687627.1687750} using Amazon’s Mechanical Turk platform, and a second subset contains queries collected by Venetis et al. \cite{Venetis:2011:RST:2002938.2002939} using  Google Squared. The second subset consists of queries created by combining the name of an ‘instance class’ and a property (i.e., attribute).
\iffalse
We present three statistics about the query collection: the
minimum length of a query is 1 term, the maximum length
of a query is 7 terms, and the average length of the queries
in the collection is 2.8 terms. %Examples of queries are: ‘world interest rates Table’, ‘fuel consumption’, ‘state capitals and
%largest cities in us’, ‘baseball teams captain’, etc. 
\fi
For the ground truth relevance scores of query-table pairs, every pair is evaluated using three numbers: 0 means irrelevant, 1 means partially relevant and 2 means relevant. %The objective of the annotators was to use the retrieved tables to create a new table that fulfills the query. %So, for a given query, they needed to find tables that are useful in forming a single table that matches the query. 
%By using this task to evaluate a given table’s relevance, if a table could not be used to create the final table, it is given a relevance 0. If only some values are used from the table, it is partially relevant. Finally, if blocks of a table are used, it is considered relevant. 
There are $60$ queries in the WikiTables collection, and $3117$ query-table pairs. 

In addition to the keyword-based table retrieval, we adapt WikiTables for the table similarity. As in TabSim \cite{tabsim}, we iterate over all the queries of WikiTables, and if two tables are relevant to a query, the table pair is given a label 1. On the other hand, an irrelevant table to a query is considered not similar to all tables that are relevant to the query, and therefore the table pair is given a label 0. 

\subsubsection{PMC}

Habibi et al.~\cite{tabsim} proposed a table corpus that is formed from PubMed Central (PMC) Open Access subset, and used for evaluation on the table similarity task. This collection is related to biomedicine and life sciences. Each table contains a caption and data values. The table pairs are annotated for binary classification by comparing the caption and data values of each table. A table pair is given a label dissimilar if both the caption and data values are labeled dissimilar, otherwise the table pair is given the label similar. In the PMC table corpus, there are $1391$ table pairs, where $542$ pairs are similar and $849$ pairs are dissimilar.

\subsubsection{Query by Example Data}

Zhang and Balog \cite{Zhang2019RecommendingRT} proposed a query by table dataset that is composed of 50 Wikipedia tables used as input queries. The query tables are related to multiple topics, and each table has at least five rows and three columns. For the ground truth relevance scores of table pairs, each pair is evaluated using three numbers: 2 means highly relevant and it indicates that the queried table is about the same topic of the query table with additional content, 1 means relevant and it indicates that the queried table contains a content that largely overlaps with the query table, and 0 means irrelevant. The total number of table pairs is $2850$.

\begin{table*}[t!]
%\scriptsize
%\small
\caption{ Table similarity results using a classification threshold equal to 0.5.
}
\begin{subtable}[t]{0.48\textwidth}
\begin{tabular}{@{}lllll@{}}
\toprule
\bf Method Name & Macro-P & Macro-R & Macro-F &Accur.  \\ \midrule
 Tfidf+MLP& 0.7834  & 0.6735  & 0.6529 & 0.6951  \\
 Embedding+MLP&0.8496  & 0.7710 &0.7736 &0.7931  \\
 Tfidf+Embedding+MLP& 0.8736& 0.8381 & 0.8447& 0.8506 \\
 TabSim \cite{tabsim}& 0.8865 & 0.8545&0.8613 & 0.8705\\
 TaBERT \cite{tabert}& 0.9109 &0.9024 &0.9055 &0.9067 \\\bottomrule
StruBERT (fine)   &0.9208   &0.9058 & 0.9104&0.9124\\
StruBERT (coarse)   &0.9276   & 0.9154&0.9194 &0.9210 \\
StruBERT (KP)   &0.9148  &0.9060 &0.9091  &0.9109 \\
StruBERT (CNN)   &0.9293 &0.9164 &0.9205  &0.9224 \\
StruBERT&\textbf{0.9321}\textsuperscript{\dag}  &\textbf{0.9284}\textsuperscript{\dag}  &\textbf{0.9300}\textsuperscript{\dag} &\textbf{0.9310}\textsuperscript{\dag} 
 \\ \bottomrule
\end{tabular}
\caption{\footnotesize PMC}
\label{tab:table1_d}
\end{subtable}
%\bigskip 
\hspace{\fill}
\begin{subtable}[t]{0.48\textwidth}
\begin{tabular}{@{}lllll@{}}
\toprule
\bf Method Name & Macro-P & Macro-R & Macro-F &Accur.  \\ \midrule
 Tfidf+MLP&0.6256   & 0.5022  & 0.3559 & 0.5378  \\
 Embedding+MLP&0.8429  &0.8419  &0.8423 & 0.8433 \\
 Tfidf+Embedding+MLP& 0.8632&0.8554  &0.8574 & 0.8594 \\
 TabSim \cite{tabsim}& 0.8480 & 0.8458&0.8466 &0.8478  \\
 TaBERT \cite{tabert}& 0.9696 &0.9626 &0.9649 &0.9653 \\\bottomrule
StruBERT (fine)   & 0.9850  &0.9852 &0.9851 &0.9852\\
StruBERT (coarse)   &0.9838   &0.9816 & 0.9825&0.9826\\
StruBERT (KP)   & 0.9733 & 0.9713&0.9722  &0.9724 \\
StruBERT (CNN)   & 0.9782&0.9737 &0.9753  &0.9756 \\
StruBERT&\textbf{0.9945}\textsuperscript{\dag}  &\textbf{0.9938}\textsuperscript{\dag}  &\textbf{0.9941}\textsuperscript{\dag} &\textbf{0.9942}\textsuperscript{\dag} 
 \\ \bottomrule
\end{tabular}
\caption{\footnotesize WikiTables}
\label{tab:table1_a}
\end{subtable}
\vspace*{-1mm}

\label{metrics}
\end{table*}

\subsection{Baselines}

\subsubsection{Keyword-based Table Retrieval}

%To evaluate the performance of our proposed method 
For evaluation on the keyword-based table retrieval, we compare against the following baselines: 

\textbf{MultiField-BM25}:
In a multi-field ranking scenario, a table is defined using multiple fields. \iffalse For example, in WikiTables data, the fields are: page title, section title, table caption, attributes and table body or values. \fi MultiField-BM25 combines BM25 \cite{Robertson96okapiat} scores for multi-field tables. 

\iffalse
\textbf{MCON}: This baseline is based on new word embeddings for attribute tokens of tables. When calculating the retrieval score of MCON, we use the MaxTable ranking method which was shown to be the best ranking method for unsupervised table retrieval~\cite{Trabelsi}. MaxTable is a late fusion similarity model that finds the closest table token to each query token
using cosine similarity, and then sums over these similarities.
\fi

\textbf{STR} \cite{DBLP:journals/corr/abs-1802-06159}: Multiple embedding-based features are computed for a table and query, then different matching strategies are used to generate the ranking features from the embeddings. A random forest is used to predict the relevance score of a query-table pair. \iffalse using the ranking features.\fi

\textbf{BERT-Row-Max} \cite{Chen2020TableSU}: The [CLS] embedding of the sequence formed from the query and table is used to predict the relevance score of a query-table pair.

\textbf{DSRMM} \cite{dsrmm}: A joint model that captures both semantic and relevance matching signals from a query-table pair to predict a real-valued relevance score.  

\textbf{TaBERT} \cite{tabert}: A %BERT-based 
 model that is originally proposed for semantic parsing over tables. We use the embedding of the [CLS] token from the last layer of the vertical self-attention as \iffalse the ranking feature. A MLP layer is then used to predict the relevance score of a given query-table pair. \fi
input to a MLP layer.

 %We do not compare to \textbf{MTR}~\cite{shraga2020web} because its source is unavailable, and thus we could not produce results according to our experimental settings, in particular with respect to train/test splits. 
 
 Due to limited computational resources, we use the BERT-base-cased %\footnote{\url{https://huggingface.co/bert-base-cased}} 
 for our method StruBERT, and for BERT-based baselines which are BERT-Row-Max and TaBERT. We note that the original paper of BERT-Row-Max uses the BERT-large-cased.

\subsubsection{Table Matching}

%To evaluate the performance of our proposed method 
For evaluation in table matching, we compare against the following baselines: 

\textbf{Embedding+MLP}: A table is flattened and concatenated with the metadata to form a single document for each table. Then, the mean of word embeddings using Glove \cite{glove} is calculated for each table. The final ranking feature is computed using pointwise multiplication between the embeddings of tables, then forwarded to a MLP layer to predict the relevance score. 

\textbf{TF-IDF+MLP}: TableRank \cite{tablerank} computes Term Frequency-Inverse Document Frequency (TF-IDF) for tables. The TF-IDF score is computed using the metadata and values of a given table, instead of the document that contains the table. A MLP layer is used instead of the cosine similarity to predict the semantic matching score. %of a given table pair.

\textbf{TabSim} \cite{tabsim}: Two separate neural network models are introduced to form the representations of a table: one model extracts the embedding from the caption, and a second model extracts column embeddings from the data values.

\textbf{TaBERT} \cite{tabert}: A TaBERT-based Siamese model is used to evaluate the semantic similarity between tables. For a given table, we extract the [CLS] embedding obtained from applying the vertical self-attention over the row-level sequences of the table. Then, pointwise multiplication is applied between the [CLS] embeddings of both tables, and the resulting vector is forwarded to a MLP layer to predict the table matching score. 

\textbf{StruBERT (KP)}: This baseline is a variation of our method that uses a kernel pooling (KP) \cite{Xiong2017EndtoEndNA}-based ranking model on top of StruBERT features. KP is the main component of strong ranking models \cite{Xiong2017EndtoEndNA,Daiwsdm}, and we adapt KP for cross-matching of fine-grained features. We construct the interaction matrices $I=\{I_{r_ir_j},I_{c_ic_j}\}$, where $I_{r_ir_j}$, and $I_{c_ic_j}$ denote the interactions of $\boldsymbol{E_r^i}$-$\boldsymbol{E_r^j}$ and $\boldsymbol{E_c^i}$-$\boldsymbol{E_c^j}$ respectively, and the elements of each matrix are computed using cosine similarity between the embeddings of the corresponding rows and columns. To summarize each interaction matrix into a fixed-length feature vector, we use KP to extract soft-match signals between different fields of $T_i$ and $T_j$. %A linear layer is used top the KP-based feature vector to a relevance score.

\iffalse
The interaction matrix (IM) is computed using cosine similarity between the embeddings of %the corresponding rows and columns 
 columns (rows) of a table pair $(T_i,T_j)$. To summarize each interaction matrix into a %fixed-length
feature vector, kernel pooling \cite{Xiong2017EndtoEndNA} is used. %to extract soft-match signals between different fields of $T_i$ and $T_j$. 
Kernel pooling is the main component of strong ranking models \cite{Xiong2017EndtoEndNA,Daiwsdm}, and we adapt kernel pooling for cross-matching of fine-grained features.  We construct the interaction matrices $M=\{M_{r_ir_j},M_{c_ic_j}\}$, where $M_{r_ir_j}$, and $M_{c_ic_j}$ denote the interactions of $\boldsymbol{E_r^i}$-$\boldsymbol{E_r^j}$, $\boldsymbol{E_c^i}$-$\boldsymbol{E_c^j}$ respectively, and the elements of each matrix are computed using cosine similarity between the embeddings of the corresponding rows and columns. To summarize each interaction matrix into a fixed-length feature vector, we use kernel pooling to extract soft-match signals between different fields of $T_i$ and $T_j$.
\fi

\textbf{StruBERT (CNN)}: This baseline is a variation of our method that uses Convolutional Neural Networks (CNN) on top of StruBERT features. This baseline is based on the interaction tensor, denoted by $\mathcal{S}$, which is computed using pointwise multiplication 
between pairwise column (row) embeddings of a table pair. Inspired by DeepRank \cite{DeepRank}, we use one layer of CNN filters with all possible combinations of widths and heights that are applied to $\mathcal{S}$: 
\begin{equation}
h_{i, j}^{(\kappa)}= \sum_{s=1}^{\gamma} \sum_{t=1}^{\gamma} \left( \sum_{l=1}^{d} w_{s, t}^{l (\kappa)} \cdot \mathcal{S}_{i:i+s, j:j+t}^{(l)}\right)+b_{s,t}^{(\kappa)}, \textrm{ } \kappa=1, \cdots, K
\end{equation}
where $\gamma$ is the maximum size of a CNN filter, $\mathcal{S}_{i:i+s, j:j+t}^{(l)}$ is a $s \times t$ matrix from the $l$-th channel of $\mathcal{S}$ starting from $i$-th row and $j$-th column, $K$ is the total number of CNN filters, and $w_{s, t}^{l (\kappa)}$ and $b_{s,t}^{(\kappa)}$ are parameters of CNN. Then,  we keep only the most significant matching signal from each feature map to form a single vector.

\subsection{Experimental Setup}

Our model is implemented using PyTorch, with two Nvidia GeForce GTX 1080. For keyword- and content-based table retrieval, the parameters of our model are updated using a mean square error pointwise loss between predicted and groundtruth relevance scores, and for table similarity, we use the cross-entropy loss function. The dimension $d$ is equal to 768. The number of Transformer layers $H$ and $V$ in the horizontal and vertical self-attention, respectively, are equal to 3. In StruBERT, BERT-base-cased and the vertical self-attention are initialized using TaBERT$_{Base}(K=3)$\footnote{https://github.com/facebookresearch/TaBERT} which is pretrained using content snapshots with 3 rows.  Such pretraining requires high-memory GPUs that are not currently possessed by our team; therefore we randomly initialize the horizontal self-attention %\footnote{We tried to initialize the horizontal self-attention using the vertical self-attention from TaBERT \cite{tabert}, but we did not notice an improvement in the reported metrics.} 
 so that the row-based dependencies are only captured during fine-tuning on the target dataset. We expect an increase in the results with pretraining the horizontal self-attention on similar task to the Masked Column Prediction (MCP) from TaBERT \cite{tabert} (in our case, the pretraining task should be a Masked Row Prediction). We leave pretraining the horizontal self-attention as a future direction. % when hardware resources are available. 

%For keyword- and content-based table retrieval, the parameters of our model are updated using a mean square error pointwise loss between the predicted and the groundtruth relevance score. For table similarity, we update the parameters of our model using cross-entropy loss function. We use the Adam \cite{adam} optimizer for gradient descent to minimize the loss functions.

\subsection{Experimental Results}
We report results using five-fold cross validation. For keyword-based table retrieval, we use the same splits as Chen et al.~\cite{Chen2020TableSU} to report results on five-fold cross validation for our method and baselines.
We evaluate the performance of our proposed method and baselines on the keyword- and content-based table retrieval tasks using Normalized Discounted Cumulative Gain (NDCG) \cite{jarvelin2002cumulated}, 
Mean Reciprocal Rank (MRR), and Mean Average Precision (MAP). %All evaluation metrics results are reported using the TREC evaluation software, trec\_eval\footnote{https://trec.nist.gov/trec\_eval/trec\_eval.8.1.tar.gz}. 
We evaluate the performance of our method and baselines on the table similarity task using macro-averaged precision (P), recall (R) and F-score, and accuracy of predictions on the testing set. To test significance,
we use the paired Student’s t-test and write $\dag$ to denote significance at the 0.05 level over
all other methods.
%The dimension $d$ is equal to 768. The number of Transformer layers $H$ and $V$ in the horizontal and vertical self-attention, respectively, are equal to 3. 
\begin{table}
\caption{Content-based table retrieval results on the query by example dataset \cite{Zhang2019RecommendingRT}.}
\label{tab:my-table}
\resizebox{0.35\textwidth}{!}{%
\begin{tabular}{@{}lccc@{}}
\toprule
Model & \textbf{NDCG@5} &\textbf{MRR}&\textbf{MAP} \\ \midrule
BM25   & 0.5369  &0.5832 &0.5417 \\ 
DSRMM \cite{dsrmm}  & 0.5768  &0.6193 &0.5914 \\
TabSim \cite{tabsim}  & 0.5739  &0.6056 &0.5932\\ 
TaBERT \cite{tabert}  & 0.5877  & 0.6120 &0.5942 \\ \bottomrule
StruBERT (fine)   &0.6015   &0.6419 &0.6091 \\
StruBERT (coarse)   &0.6140   &0.6478  & 0.6142\\
StruBERT (KP)   &0.5990  &0.6200 & 0.5959\\
StruBERT (CNN)   &0.6177 &0.6378 & 0.6179\\
StruBERT  & \textbf{0.6345}\textsuperscript{\dag}  & \textbf{0.6601}\textsuperscript{\dag} &\textbf{0.6297}\\ \bottomrule
\end{tabular}\label{content}}
\end{table}

\subsubsection{Table Similarity Results}

Table \ref{metrics}(a) shows the performance of different approaches on the PMC collection. Given that table similarity is an instance of table matching, StruBERT features are used based on the steps that are described in Section \ref{table_matching}.  We show that our proposed method StruBERT outperforms the baselines for all evaluation metrics. By incorporating the structural and textual features into a cross-matching based ranking model, we were able to capture the semantic similarity between tables both in term of tabular content and metadata, and this leads to an increase in evaluation metrics compared to baselines that either ignore the structural information or treat the textual and structural information separately. Considering the table as a single document in TF-IDF and embedding baselines lead to the lowest results, which indicates that the structural similarity between tables is an important factor in table similarity. %Figure \ref{accuracy}(a) shows the ROC results on PMC dataset  where our method
%SrtuBERT outperforms the baselines. 
The results on this dataset show a clear advantage from using embedding-based features (traditional or contextualized) compared to term frequency features that are based on exact matching. StruBERT (fine) and StruBERT (coarse) show ablation study results for predicting semantic similarity using only fine- and coarse-grained features, respectively. By combining both categories of features, we achieve higher evaluation metric results.

Table \ref{metrics}(b) %and Figure \ref{accuracy}(b) 
shows the performance of the different approaches on the WikiTables. Consistent with PMC, our results on the WikiTables show the importance of the structure- and context-aware features in improving the table similarity prediction. Table similarity results on WikiTables and PMC show that StruBERT achieves significant improvements in the evaluation metrics of two data table collections from different domains, which supports the generalization characteristic of our proposed method.

\begin{table}
%\vspace*{-5mm}
\caption{Keyword-based table retrieval results on the WikiTables dataset \cite{Bhagavatula2015TabELEL}.}
\resizebox{0.35\textwidth}{!}{%
\begin{tabular}{@{}lccc@{}}
\toprule
Model & \textbf{NDCG@5} &\textbf{MRR}&\textbf{MAP} \\ \midrule
MultiField-BM25   & 0.4365  & 0.4882&0.4596 \\ 
MCON \cite{Trabelsi}  & 0.5152  &0.5321 &0.5193  \\
STR \cite{DBLP:journals/corr/abs-1802-06159}  &0.5762   & 0.6062&0.5711 \\
DSRMM \cite{dsrmm}  &0.5978   &0.6390 & 0.5992\\
TaBERT \cite{tabert}  &0.6055   & 0.6462& 0.6123\\ 
BERT-Row-Max \cite{Chen2020TableSU} & 0.6167  & 0.6436& 0.6146\\ \bottomrule
StruBERT (fine)  & 0.6000   & 0.6406  & 0.6020\\
StruBERT (coarse)  & 0.6217   & 0.6562  & 0.6225\\
StruBERT & \textbf{0.6393}\textsuperscript{\dag}   &\textbf{0.6688}\textsuperscript{\dag}   &\textbf{0.6378}\\ \bottomrule
\end{tabular}\label{keyword_tr}}
\end{table}

\subsubsection{Content-based Table Retrieval Results}
Table \ref{content} shows the content-based table retrieval results. Given that content-based table retrieval is an instance of table matching, StruBERT features are used based on the steps that are described in Section \ref{table_matching}. The StruBERT model that combines fine- and coarse-grained features achieves a $7.96\%$ improvement in terms of NDCG@5 upon TaBERT that only uses the [CLS] embedding  obtained from the vertical self-attention. We also report ablation study results where we predict the semantic similarity between tables using only fine- or coarse-grained features. Both categories of features lead to better retrieval results than the baselines, and by combining both the fine- and coarse-grained features, we capture the textual and structural similarities between tables. An important step in table similarity assessment is the order-invariant cross-matching between columns (rows) of tables which is satisfied using miniBERT as a ranking model on top of StruBERT features. 

Our approach uses a novel miniBERT model to map StruBERT features to a relevance score. We investigate the performance of alternative ranking models when given StruBERT features. %We compare miniBERT against two ranking baselines that have been frequently used in document retrieval. We can evaluate the performance of the various ranking models by comparing all of them on the same task and StruBERT features.
Table \ref{content} shows the results of comparing miniBERT against StruBERT (KP) and StruBERT (CNN) in the case of content-based table retrieval. miniBERT outperforms both baselines in all evaluation metrics.
Kernel pooling summarizes the one-to-one matching signals computed in the interaction matrix to a single feature vector. So, StruBERT (KP) does not solve the case of many-to-many matching of rows or columns. On the other hand, by applying CNN to the interaction tensor, we capture the semantic similarity between multiple columns (rows) from a table pair, so StruBERT (CNN) captures the many-to-many matching signals. However, the convolution operation captures local interactions, and is not permutation invariant in the sense that rows and columns could be
arbitrarily ordered without changing the meaning of the table. miniBERT deals with both many-to-many matching and the permutation invariant property by taking advantage of self-attention heads.

\subsubsection{Keyword-based Table Retrieval Results}

Table \ref{keyword_tr} shows the performance of different approaches on the WikiTables collection. In keyword-based table retrieval, StruBERT features are used based on the steps that are described in Section \ref{keyword_table_retrieval}. We show that our proposed method StruBERT outperforms the baselines for all evaluation metrics. %Consistent with what has been shown in ad hoc document retrieval, supervised approaches perform better on ad hoc table retrieval than unsupervised approaches (MultiField-BM25 and MCON). 
 By adding the query to the textual information of a given table, we obtain fine- and coarse-grained features that capture early interactions between the query, and the structural and textual information of a table. %\textbf{We note that we do not compare to the hybrid variation of BERT-Row-Max \cite{Chen2020TableSU}} which combines BERT features with STR \cite{DBLP:journals/corr/abs-1802-06159} for two reasons. First, our objective is to show the current best text-based ranker for table search without including any additional hand-crafted features. Second, STR requires a wide range of features that are not available for all datasets.

\iffalse
For the ablation study of the keyword-based table retrieval, we compare row-based features, denoted by StruBERT (Rows features) and composed of the row-guided [CLS] and the row embeddings, against the columnar features denoted by StruBERT (Cols features) and composed of the column-guided [CLS] and the column embeddings. The column-based StruBERT features lead to slightly better results than the row-based StruBERT features for WikiTables. After combining the row- and column-based features, StruBERT captures the semantic similarity between the query and the textual information, and the query and the structural information defined by rows and columns.   
\fi

For the ablation study of the keyword-based table retrieval, we notice that summarizing the table and query using the [CLS] token in BERT-Row-Max, TaBERT, and StruBERT (coarse) leads to better results than fine-grained features of StruBERT (fine). This means that there are more natural language queries in the keyword-based table retrieval for WikiTables collection that are relevant to a high level summary of the textual and structural information than the specific details captured by rows and columns. After combining the fine- and coarse-grained features for all query-table pairs, StruBERT captures the semantic similarity between the query and the textual information, and the query and the structural information defined by rows and columns, and this leads to the best retrieval metrics. 

\section{Conclusions}

We have shown that a structure-aware model should augment the vertical self-attention with our novel horizontal self-attention to more accurately capture the structural information of a table, and that when we combine this with textual information using a BERT-based model, the resulting StruBERT system outperforms the state-of-the-art results in three table-related tasks: keyword- and content-based table retrieval, and table similarity. 
\iffalse
Our proposed model produces four features that can be categorized into two groups: the fine-grained features which are the contextualized row and column embeddings, and the coarse-grained features which are the row- and column-guided [CLS] embeddings.
\fi
StruBERT embeddings are integrated into our miniBERT ranking model to predict the relevance score between a keyword-based query and a table, or between a table pair. Despite being a general model, StruBERT outperforms all baselines on three different tasks, achieving a near perfect accuracy of $0.9942$ on table similarity for WikiTables, and improvement in table search for both keyword- and table-based queries with up to $7.96\%$ increase in NDCG@5 score for content-based table retrieval.  
%StruBERT outperforms the best baseline in table similarity, content-based table retrieval, and keyword-based table retrieval, achieving up to $9.96\%$, $7.96\%$, and $9.96\%$ improvement in NDCG@5 score, respectively.
%StruBERT achieves a near perfect accuracy of 0.9942 on table similarity for WikiTables. %For content-based table retrieval and keyword-based table retrieval, it scores ~0.05 and ~0.02 better on NDCG@5 than the state-of-the-art TaBERT. 
An ablation study shows that using both fine- and coarse-grained features achieves better results than either set alone. We also demonstrate that using miniBERT as a ranking model for StruBERT features outperforms other common ranking models. 
%Future work includes investigating StruBERT features in schema matching with heterogeneity where one column from one table can match several columns from a second table. 

\subsection*{Acknowledgments}
This material is based upon work supported by the National
Science Foundation under Grant No.\ IIS-1816325.

\bibliographystyle{ACM-Reference-Format}
\bibliography{ref1}

%%% -*-BibTeX-*-
%%% Do NOT edit. File created by BibTeX with style
%%% ACM-Reference-Format-Journals [18-Jan-2012].

\begin{thebibliography}{57}

%%% ====================================================================
%%% NOTE TO THE USER: you can override these defaults by providing
%%% customized versions of any of these macros before the \bibliography
%%% command.  Each of them MUST provide its own final punctuation,
%%% except for \shownote{}, \showDOI{}, and \showURL{}.  The latter two
%%% do not use final punctuation, in order to avoid confusing it with
%%% the Web address.
%%%
%%% To suppress output of a particular field, define its macro to expand
%%% to an empty string, or better, \unskip, like this:
%%%
%%% \newcommand{\showDOI}[1]{\unskip}   % LaTeX syntax
%%%
%%% \def \showDOI #1{\unskip}           % plain TeX syntax
%%%
%%% ====================================================================

\ifx \showCODEN    \undefined \def \showCODEN     #1{\unskip}     \fi
\ifx \showDOI      \undefined \def \showDOI       #1{#1}\fi
\ifx \showISBNx    \undefined \def \showISBNx     #1{\unskip}     \fi
\ifx \showISBNxiii \undefined \def \showISBNxiii  #1{\unskip}     \fi
\ifx \showISSN     \undefined \def \showISSN      #1{\unskip}     \fi
\ifx \showLCCN     \undefined \def \showLCCN      #1{\unskip}     \fi
\ifx \shownote     \undefined \def \shownote      #1{#1}          \fi
\ifx \showarticletitle \undefined \def \showarticletitle #1{#1}   \fi
\ifx \showURL      \undefined \def \showURL       {\relax}        \fi
% The following commands are used for tagged output and should be
% invisible to TeX
\providecommand\bibfield[2]{#2}
\providecommand\bibinfo[2]{#2}
\providecommand\natexlab[1]{#1}
\providecommand\showeprint[2][]{arXiv:#2}

\bibitem[\protect\citeauthoryear{Bhagavatula, Noraset, and Downey}{Bhagavatula
  et~al\mbox{.}}{2015}]%
        {Bhagavatula2015TabELEL}
\bibfield{author}{\bibinfo{person}{Chandra Bhagavatula},
  \bibinfo{person}{Thanapon Noraset}, {and} \bibinfo{person}{Doug Downey}.}
  \bibinfo{year}{2015}\natexlab{}.
\newblock \showarticletitle{TabEL: Entity Linking in Web Tables}. In
  \bibinfo{booktitle}{\emph{International Semantic Web Conference}}.
\newblock


\bibitem[\protect\citeauthoryear{Bhagavatula, Noraset, and Downey}{Bhagavatula
  et~al\mbox{.}}{2013}]%
        {Bhagavatula:2013:MEM:2501511.2501516}
\bibfield{author}{\bibinfo{person}{Chandra~Sekhar Bhagavatula},
  \bibinfo{person}{Thanapon Noraset}, {and} \bibinfo{person}{Doug Downey}.}
  \bibinfo{year}{2013}\natexlab{}.
\newblock \showarticletitle{Methods for Exploring and Mining Tables on
  Wikipedia}. In \bibinfo{booktitle}{\emph{Proceedings of the ACM SIGKDD
  Workshop on Interactive Data Exploration and Analytics}}.
  \bibinfo{publisher}{ACM}, \bibinfo{pages}{18--26}.
\newblock


\bibitem[\protect\citeauthoryear{Bromley, Bentz, Bottou, Guyon, LeCun, Moore,
  Sackinger, and Shah}{Bromley et~al\mbox{.}}{1993}]%
        {Lecun}
\bibfield{author}{\bibinfo{person}{Jane Bromley}, \bibinfo{person}{James
  Bentz}, \bibinfo{person}{Leon Bottou}, \bibinfo{person}{Isabelle Guyon},
  \bibinfo{person}{Yann LeCun}, \bibinfo{person}{Cliff Moore},
  \bibinfo{person}{Eduard Sackinger}, {and} \bibinfo{person}{Rookpak Shah}.}
  \bibinfo{year}{1993}\natexlab{}.
\newblock \showarticletitle{Signature Verification using a "Siamese" Time Delay
  Neural Network}.
\newblock \bibinfo{journal}{\emph{International Journal of Pattern Recognition
  and Artificial Intelligence}}  \bibinfo{volume}{7} (\bibinfo{date}{08}
  \bibinfo{year}{1993}), \bibinfo{pages}{25}.
\newblock


\bibitem[\protect\citeauthoryear{Cafarella, Halevy, and Khoussainova}{Cafarella
  et~al\mbox{.}}{2009}]%
        {Cafarella:2009:DIR:1687627.1687750}
\bibfield{author}{\bibinfo{person}{Michael~J. Cafarella}, \bibinfo{person}{Alon
  Halevy}, {and} \bibinfo{person}{Nodira Khoussainova}.}
  \bibinfo{year}{2009}\natexlab{}.
\newblock \showarticletitle{Data Integration for the Relational Web}.
\newblock \bibinfo{journal}{\emph{Proc. VLDB Endow.}} \bibinfo{volume}{2},
  \bibinfo{number}{1} (\bibinfo{date}{Aug.} \bibinfo{year}{2009}),
  \bibinfo{pages}{1090--1101}.
\newblock
\showISSN{2150-8097}


\bibitem[\protect\citeauthoryear{Cafarella, Halevy, Wang, Wu, and
  Zhang}{Cafarella et~al\mbox{.}}{2008}]%
        {Cafarella:2008:WEP:1453856.1453916}
\bibfield{author}{\bibinfo{person}{Michael~J. Cafarella}, \bibinfo{person}{Alon
  Halevy}, \bibinfo{person}{Daisy~Zhe Wang}, \bibinfo{person}{Eugene Wu}, {and}
  \bibinfo{person}{Yang Zhang}.} \bibinfo{year}{2008}\natexlab{}.
\newblock \showarticletitle{WebTables: Exploring the Power of Tables on the
  Web}.
\newblock \bibinfo{journal}{\emph{Proc. VLDB Endow.}} \bibinfo{volume}{1},
  \bibinfo{number}{1} (\bibinfo{date}{Aug.} \bibinfo{year}{2008}),
  \bibinfo{pages}{538--549}.
\newblock
\showISSN{2150-8097}


\bibitem[\protect\citeauthoryear{Chen, Jia, Heflin, and Davison}{Chen
  et~al\mbox{.}}{2018}]%
        {chen2018generating}
\bibfield{author}{\bibinfo{person}{Zhiyu Chen}, \bibinfo{person}{Haiyan Jia},
  \bibinfo{person}{Jeff Heflin}, {and} \bibinfo{person}{Brian~D Davison}.}
  \bibinfo{year}{2018}\natexlab{}.
\newblock \showarticletitle{Generating schema labels through dataset content
  analysis}. In \bibinfo{booktitle}{\emph{Companion Proceedings of the The Web
  Conference 2018}}. \bibinfo{pages}{1515--1522}.
\newblock


\bibitem[\protect\citeauthoryear{Chen, Jia, Heflin, and Davison}{Chen
  et~al\mbox{.}}{2020a}]%
        {chen2020leveraging}
\bibfield{author}{\bibinfo{person}{Zhiyu Chen}, \bibinfo{person}{Haiyan Jia},
  \bibinfo{person}{Jeff Heflin}, {and} \bibinfo{person}{Brian~D Davison}.}
  \bibinfo{year}{2020}\natexlab{a}.
\newblock \showarticletitle{Leveraging schema labels to enhance dataset
  search}.
\newblock \bibinfo{journal}{\emph{Advances in Information Retrieval}}
  \bibinfo{volume}{12035} (\bibinfo{year}{2020}), \bibinfo{pages}{267}.
\newblock


\bibitem[\protect\citeauthoryear{Chen, Trabelsi, Heflin, Xu, and Davison}{Chen
  et~al\mbox{.}}{2020b}]%
        {Chen2020TableSU}
\bibfield{author}{\bibinfo{person}{Zhiyu Chen}, \bibinfo{person}{Mohamed
  Trabelsi}, \bibinfo{person}{Jeff Heflin}, \bibinfo{person}{Yinan Xu}, {and}
  \bibinfo{person}{Brian~D. Davison}.} \bibinfo{year}{2020}\natexlab{b}.
\newblock \showarticletitle{Table Search Using a Deep Contextualized Language
  Model}. In \bibinfo{booktitle}{\emph{Proceedings of the 43rd International
  ACM SIGIR Conference on Research and Development in Information Retrieval}}.
  \bibinfo{publisher}{Association for Computing Machinery},
  \bibinfo{pages}{589–598}.
\newblock


\bibitem[\protect\citeauthoryear{Chen, Trabelsi, Heflin, Yin, and Davison}{Chen
  et~al\mbox{.}}{2021a}]%
        {chen2021mgnets}
\bibfield{author}{\bibinfo{person}{Zhiyu Chen}, \bibinfo{person}{Mohamed
  Trabelsi}, \bibinfo{person}{Jeff Heflin}, \bibinfo{person}{Dawei Yin}, {and}
  \bibinfo{person}{Brian~D Davison}.} \bibinfo{year}{2021}\natexlab{a}.
\newblock \showarticletitle{MGNETS: Multi-Graph Neural Networks for Table
  Search}. In \bibinfo{booktitle}{\emph{Proceedings of the 30th ACM
  International Conference on Information \& Knowledge Management}}.
  \bibinfo{pages}{2945--2949}.
\newblock


\bibitem[\protect\citeauthoryear{Chen, Zhang, and Davison}{Chen
  et~al\mbox{.}}{2021b}]%
        {chen2021wtr}
\bibfield{author}{\bibinfo{person}{Zhiyu Chen}, \bibinfo{person}{Shuo Zhang},
  {and} \bibinfo{person}{Brian~D Davison}.} \bibinfo{year}{2021}\natexlab{b}.
\newblock \showarticletitle{WTR: A Test Collection for Web Table Retrieval}.
\newblock \bibinfo{journal}{\emph{arXiv preprint arXiv:2105.02354}}
  (\bibinfo{year}{2021}).
\newblock


\bibitem[\protect\citeauthoryear{Crestan and Pantel}{Crestan and
  Pantel}{2010}]%
        {Crestan2010AFT}
\bibfield{author}{\bibinfo{person}{Eric Crestan} {and} \bibinfo{person}{Patrick
  Pantel}.} \bibinfo{year}{2010}\natexlab{}.
\newblock \showarticletitle{A fine-grained taxonomy of tables on the web}. In
  \bibinfo{booktitle}{\emph{Proceedings of the 19th {ACM} Conference on
  Information and Knowledge Management, {CIKM} 2010,}}.
  \bibinfo{publisher}{{ACM}}, \bibinfo{pages}{1405--1408}.
\newblock


\bibitem[\protect\citeauthoryear{Crestan and Pantel}{Crestan and
  Pantel}{2011}]%
        {crestan2011}
\bibfield{author}{\bibinfo{person}{Eric Crestan} {and} \bibinfo{person}{Patrick
  Pantel}.} \bibinfo{year}{2011}\natexlab{}.
\newblock \showarticletitle{Web-scale table census and classification}. In
  \bibinfo{booktitle}{\emph{Proceedings of the Forth International Conference
  on Web Search and Web Data Mining, {WSDM} 2011}}. \bibinfo{publisher}{{ACM}},
  \bibinfo{pages}{545--554}.
\newblock


\bibitem[\protect\citeauthoryear{Dai and Callan}{Dai and Callan}{2019}]%
        {dai2019}
\bibfield{author}{\bibinfo{person}{Zhuyun Dai} {and} \bibinfo{person}{Jamie
  Callan}.} \bibinfo{year}{2019}\natexlab{}.
\newblock \showarticletitle{Deeper Text Understanding for IR with Contextual
  Neural Language Modeling}. In \bibinfo{booktitle}{\emph{Proceedings of the
  42nd International ACM SIGIR Conference on Research and Development in
  Information Retrieval}}.
\newblock


\bibitem[\protect\citeauthoryear{Dai, Xiong, Callan, and Liu}{Dai
  et~al\mbox{.}}{2018}]%
        {Daiwsdm}
\bibfield{author}{\bibinfo{person}{Zhuyun Dai}, \bibinfo{person}{Chenyan
  Xiong}, \bibinfo{person}{Jamie Callan}, {and} \bibinfo{person}{Zhiyuan Liu}.}
  \bibinfo{year}{2018}\natexlab{}.
\newblock \showarticletitle{Convolutional Neural Networks for Soft-Matching
  N-Grams in Ad-Hoc Search}. In \bibinfo{booktitle}{\emph{Proceedings of the
  Eleventh ACM International Conference on Web Search and Data Mining}}.
  \bibinfo{pages}{126–134}.
\newblock


\bibitem[\protect\citeauthoryear{Das~Sarma, Fang, Gupta, Halevy, Lee, Wu, Xin,
  and Yu}{Das~Sarma et~al\mbox{.}}{2012}]%
        {DasSarma:2012:FRT:2213836.2213962}
\bibfield{author}{\bibinfo{person}{Anish Das~Sarma}, \bibinfo{person}{Lujun
  Fang}, \bibinfo{person}{Nitin Gupta}, \bibinfo{person}{Alon Halevy},
  \bibinfo{person}{Hongrae Lee}, \bibinfo{person}{Fei Wu},
  \bibinfo{person}{Reynold Xin}, {and} \bibinfo{person}{Cong Yu}.}
  \bibinfo{year}{2012}\natexlab{}.
\newblock \showarticletitle{Finding Related Tables}. In
  \bibinfo{booktitle}{\emph{Proceedings of the 2012 ACM SIGMOD International
  Conference on Management of Data}}. \bibinfo{pages}{817--828}.
\newblock


\bibitem[\protect\citeauthoryear{Devlin, Chang, Lee, and Toutanova}{Devlin
  et~al\mbox{.}}{2019}]%
        {Devlin2019BERTPO}
\bibfield{author}{\bibinfo{person}{Jacob Devlin}, \bibinfo{person}{Ming-Wei
  Chang}, \bibinfo{person}{Kenton Lee}, {and} \bibinfo{person}{Kristina
  Toutanova}.} \bibinfo{year}{2019}\natexlab{}.
\newblock \showarticletitle{BERT: Pre-training of Deep Bidirectional
  Transformers for Language Understanding}. In
  \bibinfo{booktitle}{\emph{NAACL-HLT}}.
\newblock


\bibitem[\protect\citeauthoryear{Gonzalez, Halevy, Jensen, Langen, Madhavan,
  Shapley, and Shen}{Gonzalez et~al\mbox{.}}{2010}]%
        {table_fusion}
\bibfield{author}{\bibinfo{person}{Hector Gonzalez}, \bibinfo{person}{Alon~Y.
  Halevy}, \bibinfo{person}{Christian~S. Jensen}, \bibinfo{person}{Anno
  Langen}, \bibinfo{person}{Jayant Madhavan}, \bibinfo{person}{Rebecca
  Shapley}, {and} \bibinfo{person}{Warren Shen}.}
  \bibinfo{year}{2010}\natexlab{}.
\newblock \showarticletitle{Google fusion tables: data management, integration
  and collaboration in the cloud}. In \bibinfo{booktitle}{\emph{Proceedings of
  the 1st {ACM} Symposium on Cloud Computing, SoCC 2010}}.
  \bibinfo{publisher}{{ACM}}, \bibinfo{pages}{175--180}.
\newblock


\bibitem[\protect\citeauthoryear{Guo, Fan, Ai, and Croft}{Guo
  et~al\mbox{.}}{2016}]%
        {Guo2016ADR}
\bibfield{author}{\bibinfo{person}{Jiafeng Guo}, \bibinfo{person}{Yixing Fan},
  \bibinfo{person}{Qingyao Ai}, {and} \bibinfo{person}{W.~Bruce Croft}.}
  \bibinfo{year}{2016}\natexlab{}.
\newblock \showarticletitle{A Deep Relevance Matching Model for Ad-hoc
  Retrieval}. In \bibinfo{booktitle}{\emph{Proceedings of the 25th {ACM}
  International Conference on Information and Knowledge Management, {CIKM}
  2016}}. \bibinfo{publisher}{{ACM}}, \bibinfo{pages}{55--64}.
\newblock


\bibitem[\protect\citeauthoryear{Habibi, Starlinger, and Leser}{Habibi
  et~al\mbox{.}}{2020}]%
        {tabsim}
\bibfield{author}{\bibinfo{person}{Maryam Habibi}, \bibinfo{person}{Johannes
  Starlinger}, {and} \bibinfo{person}{Ulf Leser}.}
  \bibinfo{year}{2020}\natexlab{}.
\newblock \showarticletitle{TabSim: {A} Siamese Neural Network for Accurate
  Estimation of Table Similarity}.
\newblock \bibinfo{journal}{\emph{CoRR}}  \bibinfo{volume}{abs/2008.10856}
  (\bibinfo{year}{2020}).
\newblock
\showeprint[arxiv]{2008.10856}
\urldef\tempurl%
\url{https://arxiv.org/abs/2008.10856}
\showURL{%
\tempurl}


\bibitem[\protect\citeauthoryear{Hu, Lu, Li, and Chen}{Hu
  et~al\mbox{.}}{2014}]%
        {Hu2014ConvolutionalNN}
\bibfield{author}{\bibinfo{person}{Baotian Hu}, \bibinfo{person}{Zhengdong Lu},
  \bibinfo{person}{Hang Li}, {and} \bibinfo{person}{Qingcai Chen}.}
  \bibinfo{year}{2014}\natexlab{}.
\newblock \showarticletitle{Convolutional Neural Network Architectures for
  Matching Natural Language Sentences}. In \bibinfo{booktitle}{\emph{Advances
  in Neural Information Processing Systems 27: Annual Conference on Neural
  Information Processing Systems 2014}}. \bibinfo{pages}{2042--2050}.
\newblock


\bibitem[\protect\citeauthoryear{J{\"a}rvelin and
  Kek{\"a}l{\"a}inen}{J{\"a}rvelin and Kek{\"a}l{\"a}inen}{2002}]%
        {jarvelin2002cumulated}
\bibfield{author}{\bibinfo{person}{Kalervo J{\"a}rvelin} {and}
  \bibinfo{person}{Jaana Kek{\"a}l{\"a}inen}.} \bibinfo{year}{2002}\natexlab{}.
\newblock \showarticletitle{Cumulated gain-based evaluation of IR techniques}.
\newblock \bibinfo{journal}{\emph{ACM Transactions on Information Systems
  (TOIS)}} \bibinfo{volume}{20}, \bibinfo{number}{4} (\bibinfo{year}{2002}),
  \bibinfo{pages}{422--446}.
\newblock


\bibitem[\protect\citeauthoryear{Li, Shah, and Fang}{Li et~al\mbox{.}}{2016}]%
        {table_classification}
\bibfield{author}{\bibinfo{person}{Quanzhi Li}, \bibinfo{person}{Sameena Shah},
  {and} \bibinfo{person}{Rui Fang}.} \bibinfo{year}{2016}\natexlab{}.
\newblock \showarticletitle{Table classification using both structure and
  content information: {A} case study of financial documents}. In
  \bibinfo{booktitle}{\emph{2016 {IEEE} International Conference on Big Data,
  BigData 2016, Washington DC, USA, December 5-8, 2016}}.
  \bibinfo{publisher}{{IEEE} Computer Society}, \bibinfo{pages}{1778--1783}.
\newblock


\bibitem[\protect\citeauthoryear{Liu, He, Chen, and Gao}{Liu
  et~al\mbox{.}}{2019a}]%
        {liu-etal-2019-multi}
\bibfield{author}{\bibinfo{person}{Xiaodong Liu}, \bibinfo{person}{Pengcheng
  He}, \bibinfo{person}{Weizhu Chen}, {and} \bibinfo{person}{Jianfeng Gao}.}
  \bibinfo{year}{2019}\natexlab{a}.
\newblock \showarticletitle{Multi-Task Deep Neural Networks for Natural
  Language Understanding}. In \bibinfo{booktitle}{\emph{Proceedings of the 57th
  Annual Meeting of the Association for Computational Linguistics}}.
  \bibinfo{pages}{4487--4496}.
\newblock


\bibitem[\protect\citeauthoryear{Liu, Bai, Mitra, and Giles}{Liu
  et~al\mbox{.}}{2007}]%
        {tablerank}
\bibfield{author}{\bibinfo{person}{Ying Liu}, \bibinfo{person}{Kun Bai},
  \bibinfo{person}{Prasenjit Mitra}, {and} \bibinfo{person}{C.~Lee Giles}.}
  \bibinfo{year}{2007}\natexlab{}.
\newblock \showarticletitle{TableRank: {A} Ranking Algorithm for Table Search
  and Retrieval}. In \bibinfo{booktitle}{\emph{Proceedings of the Twenty-Second
  {AAAI} Conference on Artificial Intelligence, July 22-26, 2007, Vancouver,
  British Columbia, Canada}}. \bibinfo{publisher}{{AAAI} Press},
  \bibinfo{pages}{317--322}.
\newblock


\bibitem[\protect\citeauthoryear{Liu, Ott, Goyal, Du, Joshi, Chen, Levy, Lewis,
  Zettlemoyer, and Stoyanov}{Liu et~al\mbox{.}}{2019b}]%
        {Liu2019RoBERTaAR}
\bibfield{author}{\bibinfo{person}{Yinhan Liu}, \bibinfo{person}{Myle Ott},
  \bibinfo{person}{Naman Goyal}, \bibinfo{person}{Jingfei Du},
  \bibinfo{person}{Mandar Joshi}, \bibinfo{person}{Danqi Chen},
  \bibinfo{person}{Omer Levy}, \bibinfo{person}{Mike Lewis},
  \bibinfo{person}{Luke Zettlemoyer}, {and} \bibinfo{person}{Veselin
  Stoyanov}.} \bibinfo{year}{2019}\natexlab{b}.
\newblock \showarticletitle{RoBERTa: A Robustly Optimized BERT Pretraining
  Approach}.
\newblock \bibinfo{journal}{\emph{ArXiv}}  \bibinfo{volume}{abs/1907.11692}
  (\bibinfo{year}{2019}).
\newblock


\bibitem[\protect\citeauthoryear{Mu\~{n}oz, Hogan, and Mileo}{Mu\~{n}oz
  et~al\mbox{.}}{2014}]%
        {Munoz:2014:ULD:2556195.2556266}
\bibfield{author}{\bibinfo{person}{Emir Mu\~{n}oz}, \bibinfo{person}{Aidan
  Hogan}, {and} \bibinfo{person}{Alessandra Mileo}.}
  \bibinfo{year}{2014}\natexlab{}.
\newblock \showarticletitle{Using Linked Data to Mine RDF from Wikipedia's
  Tables}. In \bibinfo{booktitle}{\emph{Proceedings of the 7th ACM
  International Conference on Web Search and Data Mining, WSDM '14}}.
  \bibinfo{pages}{533--542}.
\newblock


\bibitem[\protect\citeauthoryear{Nguyen, Viet Hung~Nguyen, Weidlich, and
  Aberer}{Nguyen et~al\mbox{.}}{2015}]%
        {nguyen2015}
\bibfield{author}{\bibinfo{person}{Tam Nguyen}, \bibinfo{person}{Quoc Viet
  Hung~Nguyen}, \bibinfo{person}{Matthias Weidlich}, {and}
  \bibinfo{person}{Karl Aberer}.} \bibinfo{year}{2015}\natexlab{}.
\newblock \showarticletitle{Result selection and summarization for Web Table
  search}.
\newblock \bibinfo{journal}{\emph{Proceedings - International Conference on
  Data Engineering}}  \bibinfo{volume}{2015} (\bibinfo{date}{05}
  \bibinfo{year}{2015}), \bibinfo{pages}{231--242}.
\newblock


\bibitem[\protect\citeauthoryear{Nie, Li, and Nie}{Nie et~al\mbox{.}}{2018}]%
        {Nie2018EmpiricalSO}
\bibfield{author}{\bibinfo{person}{Yifan Nie}, \bibinfo{person}{Yanling Li},
  {and} \bibinfo{person}{Jian{-}Yun Nie}.} \bibinfo{year}{2018}\natexlab{}.
\newblock \showarticletitle{Empirical Study of Multi-level Convolution Models
  for {IR} Based on Representations and Interactions}. In
  \bibinfo{booktitle}{\emph{Proceedings of the 2018 {ACM} {SIGIR} International
  Conference on Theory of Information Retrieval, {ICTIR}}}.
  \bibinfo{publisher}{{ACM}}, \bibinfo{pages}{59--66}.
\newblock


\bibitem[\protect\citeauthoryear{Nogueira and Cho}{Nogueira and Cho}{2019}]%
        {Nogueira2019PassageRW}
\bibfield{author}{\bibinfo{person}{Rodrigo Nogueira} {and}
  \bibinfo{person}{Kyunghyun Cho}.} \bibinfo{year}{2019}\natexlab{}.
\newblock \showarticletitle{Passage Re-ranking with BERT}.
\newblock \bibinfo{journal}{\emph{ArXiv}}  \bibinfo{volume}{abs/1901.04085}
  (\bibinfo{year}{2019}).
\newblock


\bibitem[\protect\citeauthoryear{Nogueira, Yang, Cho, and Lin}{Nogueira
  et~al\mbox{.}}{2019}]%
        {Nogueira2019MultiStageDR}
\bibfield{author}{\bibinfo{person}{Rodrigo Nogueira}, \bibinfo{person}{Wei
  Yang}, \bibinfo{person}{Kyunghyun Cho}, {and} \bibinfo{person}{Jimmy Lin}.}
  \bibinfo{year}{2019}\natexlab{}.
\newblock \showarticletitle{Multi-Stage Document Ranking with BERT}.
\newblock \bibinfo{journal}{\emph{ArXiv}}  \bibinfo{volume}{abs/1910.14424}
  (\bibinfo{year}{2019}).
\newblock


\bibitem[\protect\citeauthoryear{Pang, Lan, Guo, Xu, Xu, and Cheng}{Pang
  et~al\mbox{.}}{2017}]%
        {DeepRank}
\bibfield{author}{\bibinfo{person}{Liang Pang}, \bibinfo{person}{Yanyan Lan},
  \bibinfo{person}{Jiafeng Guo}, \bibinfo{person}{Jun Xu},
  \bibinfo{person}{Jingfang Xu}, {and} \bibinfo{person}{Xueqi Cheng}.}
  \bibinfo{year}{2017}\natexlab{}.
\newblock \showarticletitle{DeepRank: A New Deep Architecture for Relevance
  Ranking in Information Retrieval}. In \bibinfo{booktitle}{\emph{Proceedings
  of the 2017 ACM on Conference on Information and Knowledge Management}}.
  \bibinfo{pages}{257–266}.
\newblock


\bibitem[\protect\citeauthoryear{Pennington, Socher, and Manning}{Pennington
  et~al\mbox{.}}{2014}]%
        {glove}
\bibfield{author}{\bibinfo{person}{Jeffrey Pennington},
  \bibinfo{person}{Richard Socher}, {and} \bibinfo{person}{Christopher
  Manning}.} \bibinfo{year}{2014}\natexlab{}.
\newblock \showarticletitle{{G}lo{V}e: Global Vectors for Word Representation}.
  In \bibinfo{booktitle}{\emph{Proceedings of the 2014 Conference on Empirical
  Methods in Natural Language Processing ({EMNLP})}}.
  \bibinfo{publisher}{Association for Computational Linguistics},
  \bibinfo{pages}{1532--1543}.
\newblock


\bibitem[\protect\citeauthoryear{Pimplikar and Sarawagi}{Pimplikar and
  Sarawagi}{2012}]%
        {Pimplikar:2012:ATQ:2336664.2336665}
\bibfield{author}{\bibinfo{person}{Rakesh Pimplikar} {and}
  \bibinfo{person}{Sunita Sarawagi}.} \bibinfo{year}{2012}\natexlab{}.
\newblock \showarticletitle{Answering Table Queries on the Web Using Column
  Keywords}.
\newblock \bibinfo{journal}{\emph{Proc. VLDB Endow.}} \bibinfo{volume}{5},
  \bibinfo{number}{10} (\bibinfo{date}{June} \bibinfo{year}{2012}),
  \bibinfo{pages}{908--919}.
\newblock
\showISSN{2150-8097}


\bibitem[\protect\citeauthoryear{Robertson, Walker, Jones, Hancock{-}Beaulieu,
  and Gatford}{Robertson et~al\mbox{.}}{1994}]%
        {Robertson96okapiat}
\bibfield{author}{\bibinfo{person}{Stephen~E. Robertson},
  \bibinfo{person}{Steve Walker}, \bibinfo{person}{Susan Jones},
  \bibinfo{person}{Micheline Hancock{-}Beaulieu}, {and} \bibinfo{person}{Mike
  Gatford}.} \bibinfo{year}{1994}\natexlab{}.
\newblock \showarticletitle{Okapi at {TREC-3}}. In
  \bibinfo{booktitle}{\emph{Proceedings of The Third Text REtrieval Conference,
  {TREC} 1994}}, Vol.~\bibinfo{volume}{500-225}. \bibinfo{pages}{109--126}.
\newblock


\bibitem[\protect\citeauthoryear{Sakata, Shibata, Tanaka, and Kurohashi}{Sakata
  et~al\mbox{.}}{2019}]%
        {sakata2019}
\bibfield{author}{\bibinfo{person}{Wataru Sakata}, \bibinfo{person}{Tomohide
  Shibata}, \bibinfo{person}{Ribeka Tanaka}, {and} \bibinfo{person}{Sadao
  Kurohashi}.} \bibinfo{year}{2019}\natexlab{}.
\newblock \showarticletitle{FAQ Retrieval Using Query-Question Similarity and
  BERT-Based Query-Answer Relevance}. In \bibinfo{booktitle}{\emph{Proceedings
  of the 42nd International ACM SIGIR Conference on Research and Development in
  Information Retrieval}}. \bibinfo{publisher}{Association for Computing
  Machinery}, \bibinfo{pages}{1113–1116}.
\newblock
\showISBNx{9781450361729}


\bibitem[\protect\citeauthoryear{Shraga, Roitman, Feigenblat, and Canim}{Shraga
  et~al\mbox{.}}{2020a}]%
        {shraga}
\bibfield{author}{\bibinfo{person}{Roee Shraga}, \bibinfo{person}{Haggai
  Roitman}, \bibinfo{person}{Guy Feigenblat}, {and} \bibinfo{person}{Mustafa
  Canim}.} \bibinfo{year}{2020}\natexlab{a}.
\newblock \showarticletitle{Ad Hoc Table Retrieval using Intrinsic and
  Extrinsic Similarities}. In \bibinfo{booktitle}{\emph{{WWW} '20: The Web
  Conference, 2020}}, \bibfield{editor}{\bibinfo{person}{Yennun Huang},
  \bibinfo{person}{Irwin King}, \bibinfo{person}{Tie{-}Yan Liu}, {and}
  \bibinfo{person}{Maarten van Steen}} (Eds.). \bibinfo{publisher}{{ACM} /
  {IW3C2}}, \bibinfo{pages}{2479--2485}.
\newblock


\bibitem[\protect\citeauthoryear{Shraga, Roitman, Feigenblat, and
  Cannim}{Shraga et~al\mbox{.}}{2020b}]%
        {shraga2020web}
\bibfield{author}{\bibinfo{person}{Roee Shraga}, \bibinfo{person}{Haggai
  Roitman}, \bibinfo{person}{Guy Feigenblat}, {and} \bibinfo{person}{Mustafa
  Cannim}.} \bibinfo{year}{2020}\natexlab{b}.
\newblock \showarticletitle{Web Table Retrieval using Multimodal Deep
  Learning}. In \bibinfo{booktitle}{\emph{Proceedings of the 43rd International
  ACM SIGIR Conference on Research and Development in Information Retrieval}}.
  \bibinfo{pages}{1399--1408}.
\newblock


\bibitem[\protect\citeauthoryear{Trabelsi, Cao, and Heflin}{Trabelsi
  et~al\mbox{.}}{2020a}]%
        {selab_arxiv}
\bibfield{author}{\bibinfo{person}{Mohamed Trabelsi}, \bibinfo{person}{Jin
  Cao}, {and} \bibinfo{person}{Jeff Heflin}.} \bibinfo{year}{2020}\natexlab{a}.
\newblock \showarticletitle{Semantic Labeling Using a Deep Contextualized
  Language Model}.
\newblock \bibinfo{journal}{\emph{CoRR}}  \bibinfo{volume}{abs/2010.16037}
  (\bibinfo{year}{2020}).
\newblock


\bibitem[\protect\citeauthoryear{Trabelsi, Cao, and Heflin}{Trabelsi
  et~al\mbox{.}}{2021a}]%
        {selab_ijcnn}
\bibfield{author}{\bibinfo{person}{Mohamed Trabelsi}, \bibinfo{person}{Jin
  Cao}, {and} \bibinfo{person}{Jeff Heflin}.} \bibinfo{year}{2021}\natexlab{a}.
\newblock \showarticletitle{SeLaB: Semantic Labeling with {BERT}}. In
  \bibinfo{booktitle}{\emph{International Joint Conference on Neural Networks,
  {IJCNN} 2021, Shenzhen, China, July 18-22, 2021}}.
  \bibinfo{publisher}{{IEEE}}, \bibinfo{pages}{1--8}.
\newblock


\bibitem[\protect\citeauthoryear{Trabelsi, Chen, Davison, and Heflin}{Trabelsi
  et~al\mbox{.}}{2020b}]%
        {dsrmm}
\bibfield{author}{\bibinfo{person}{Mohamed Trabelsi}, \bibinfo{person}{Zhiyu
  Chen}, \bibinfo{person}{Brian~D. Davison}, {and} \bibinfo{person}{Jeff
  Heflin}.} \bibinfo{year}{2020}\natexlab{b}.
\newblock \showarticletitle{A Hybrid Deep Model for Learning to Rank Data
  Tables}. In \bibinfo{booktitle}{\emph{2020 IEEE International Conference on
  Big Data (Big Data)}}.
\newblock


\bibitem[\protect\citeauthoryear{Trabelsi, Chen, Davison, and Heflin}{Trabelsi
  et~al\mbox{.}}{2020c}]%
        {multiemrgcn}
\bibfield{author}{\bibinfo{person}{Mohamed Trabelsi}, \bibinfo{person}{Zhiyu
  Chen}, \bibinfo{person}{Brian~D. Davison}, {and} \bibinfo{person}{Jeff
  Heflin}.} \bibinfo{year}{2020}\natexlab{c}.
\newblock \showarticletitle{Relational Graph Embeddings for Table Retrieval}.
  In \bibinfo{booktitle}{\emph{{IEEE} International Conference on Big Data, Big
  Data 2020}}. \bibinfo{publisher}{{IEEE}}, \bibinfo{pages}{3005--3014}.
\newblock


\bibitem[\protect\citeauthoryear{Trabelsi, Chen, Davison, and Heflin}{Trabelsi
  et~al\mbox{.}}{2021b}]%
        {survey_doc_retrieval}
\bibfield{author}{\bibinfo{person}{Mohamed Trabelsi}, \bibinfo{person}{Zhiyu
  Chen}, \bibinfo{person}{Brian~D. Davison}, {and} \bibinfo{person}{Jeff
  Heflin}.} \bibinfo{year}{2021}\natexlab{b}.
\newblock \showarticletitle{Neural ranking models for document retrieval}.
\newblock \bibinfo{journal}{\emph{Inf. Retr. J.}} \bibinfo{volume}{24},
  \bibinfo{number}{6} (\bibinfo{year}{2021}), \bibinfo{pages}{400--444}.
\newblock


\bibitem[\protect\citeauthoryear{Trabelsi, Davison, and Heflin}{Trabelsi
  et~al\mbox{.}}{2019}]%
        {Trabelsi}
\bibfield{author}{\bibinfo{person}{Mohamed Trabelsi}, \bibinfo{person}{Brian~D.
  Davison}, {and} \bibinfo{person}{Jeff Heflin}.}
  \bibinfo{year}{2019}\natexlab{}.
\newblock \showarticletitle{Improved Table Retrieval Using Multiple Context
  Embeddings for Attributes}. In \bibinfo{booktitle}{\emph{2019 IEEE
  International Conference on Big Data (Big Data)}}.
  \bibinfo{pages}{1238--1244}.
\newblock


\bibitem[\protect\citeauthoryear{Vaswani, Shazeer, Parmar, Uszkoreit, Jones,
  Gomez, Kaiser, and Polosukhin}{Vaswani et~al\mbox{.}}{2017}]%
        {transformer}
\bibfield{author}{\bibinfo{person}{Ashish Vaswani}, \bibinfo{person}{Noam
  Shazeer}, \bibinfo{person}{Niki Parmar}, \bibinfo{person}{Jakob Uszkoreit},
  \bibinfo{person}{Llion Jones}, \bibinfo{person}{Aidan~N Gomez},
  \bibinfo{person}{\L~ukasz Kaiser}, {and} \bibinfo{person}{Illia Polosukhin}.}
  \bibinfo{year}{2017}\natexlab{}.
\newblock \showarticletitle{Attention is All you Need}.
\newblock In \bibinfo{booktitle}{\emph{Advances in Neural Information
  Processing Systems 30}}. \bibinfo{pages}{5998--6008}.
\newblock


\bibitem[\protect\citeauthoryear{Wang, Singh, Michael, Hill, Levy, and
  Bowman}{Wang et~al\mbox{.}}{2018}]%
        {wang-etal-2018-glue}
\bibfield{author}{\bibinfo{person}{Alex Wang}, \bibinfo{person}{Amanpreet
  Singh}, \bibinfo{person}{Julian Michael}, \bibinfo{person}{Felix Hill},
  \bibinfo{person}{Omer Levy}, {and} \bibinfo{person}{Samuel Bowman}.}
  \bibinfo{year}{2018}\natexlab{}.
\newblock \showarticletitle{{GLUE}: A Multi-Task Benchmark and Analysis
  Platform for Natural Language Understanding}. In
  \bibinfo{booktitle}{\emph{Proceedings of the 2018 {EMNLP} Workshop
  {B}lackbox{NLP}: Analyzing and Interpreting Neural Networks for {NLP}}}.
  \bibinfo{pages}{353--355}.
\newblock


\bibitem[\protect\citeauthoryear{Wang, Sun, Chen, Pujara, and Szekely}{Wang
  et~al\mbox{.}}{2021}]%
        {gtr}
\bibfield{author}{\bibinfo{person}{Fei Wang}, \bibinfo{person}{Kexuan Sun},
  \bibinfo{person}{Muhao Chen}, \bibinfo{person}{Jay Pujara}, {and}
  \bibinfo{person}{Pedro~A. Szekely}.} \bibinfo{year}{2021}\natexlab{}.
\newblock \showarticletitle{Retrieving Complex Tables with Multi-Granular Graph
  Representation Learning}. In \bibinfo{booktitle}{\emph{The 44th International
  {ACM} {SIGIR} Conference on Research and Development in Information
  Retrieval}}. \bibinfo{publisher}{{ACM}}, \bibinfo{pages}{1472--1482}.
\newblock


\bibitem[\protect\citeauthoryear{Xiong, Dai, Callan, Liu, and Power}{Xiong
  et~al\mbox{.}}{2017}]%
        {Xiong2017EndtoEndNA}
\bibfield{author}{\bibinfo{person}{Chenyan Xiong}, \bibinfo{person}{Zhuyun
  Dai}, \bibinfo{person}{Jamie Callan}, \bibinfo{person}{Zhiyuan Liu}, {and}
  \bibinfo{person}{Russell Power}.} \bibinfo{year}{2017}\natexlab{}.
\newblock \showarticletitle{End-to-End Neural Ad-hoc Ranking with Kernel
  Pooling}. In \bibinfo{booktitle}{\emph{Proceedings of the 40th International
  ACM SIGIR Conference on Research and Development in Information Retrieval,
  SIGIR '17}}. \bibinfo{publisher}{ACM}, \bibinfo{pages}{55--64}.
\newblock


\bibitem[\protect\citeauthoryear{Yang, Zhang, and Lin}{Yang
  et~al\mbox{.}}{2019}]%
        {Yang2019SimpleAO}
\bibfield{author}{\bibinfo{person}{Wei Yang}, \bibinfo{person}{Haotian Zhang},
  {and} \bibinfo{person}{Jimmy Lin}.} \bibinfo{year}{2019}\natexlab{}.
\newblock \showarticletitle{Simple Applications of BERT for Ad Hoc Document
  Retrieval}.
\newblock \bibinfo{journal}{\emph{ArXiv}}  \bibinfo{volume}{abs/1903.10972}
  (\bibinfo{year}{2019}).
\newblock


\bibitem[\protect\citeauthoryear{Yi, Chen, Heflin, and Davison}{Yi
  et~al\mbox{.}}{2018}]%
        {yi2018recognizing}
\bibfield{author}{\bibinfo{person}{Yang Yi}, \bibinfo{person}{Zhiyu Chen},
  \bibinfo{person}{Jeff Heflin}, {and} \bibinfo{person}{Brian~D Davison}.}
  \bibinfo{year}{2018}\natexlab{}.
\newblock \showarticletitle{Recognizing quantity names for tabular data}. In
  \bibinfo{booktitle}{\emph{ProfS/KG4IR/Data: Search@ SIGIR}}.
\newblock


\bibitem[\protect\citeauthoryear{Yilmaz, Wang, Yang, Zhang, and Lin}{Yilmaz
  et~al\mbox{.}}{2019}]%
        {Yilmaz2019ApplyingBT}
\bibfield{author}{\bibinfo{person}{Zeynep~Akkalyoncu Yilmaz},
  \bibinfo{person}{Shengjin Wang}, \bibinfo{person}{Wei Yang},
  \bibinfo{person}{Haotian Zhang}, {and} \bibinfo{person}{Jimmy Lin}.}
  \bibinfo{year}{2019}\natexlab{}.
\newblock \showarticletitle{Applying {BERT} to Document Retrieval with Birch}.
  In \bibinfo{booktitle}{\emph{Proceedings of the 2019 Conference on Empirical
  Methods in Natural Language Processing and the 9th International Joint
  Conference on Natural Language Processing, {EMNLP-IJCNLP} 2019}}.
  \bibinfo{publisher}{Association for Computational Linguistics},
  \bibinfo{pages}{19--24}.
\newblock


\bibitem[\protect\citeauthoryear{Yin, Neubig, Yih, and Riedel}{Yin
  et~al\mbox{.}}{2020}]%
        {tabert}
\bibfield{author}{\bibinfo{person}{Pengcheng Yin}, \bibinfo{person}{Graham
  Neubig}, \bibinfo{person}{Wen-tau Yih}, {and} \bibinfo{person}{Sebastian
  Riedel}.} \bibinfo{year}{2020}\natexlab{}.
\newblock \showarticletitle{{T}a{BERT}: Pretraining for Joint Understanding of
  Textual and Tabular Data}. In \bibinfo{booktitle}{\emph{Proceedings of the
  58th Annual Meeting of the Association for Computational Linguistics}}.
  \bibinfo{publisher}{Association for Computational Linguistics},
  \bibinfo{pages}{8413--8426}.
\newblock


\bibitem[\protect\citeauthoryear{Yoshida and Torisawa}{Yoshida and
  Torisawa}{2001}]%
        {Yoshida01amethod}
\bibfield{author}{\bibinfo{person}{Minoru Yoshida} {and}
  \bibinfo{person}{Kentaro Torisawa}.} \bibinfo{year}{2001}\natexlab{}.
\newblock \showarticletitle{A method to integrate tables of the World Wide
  Web}. In \bibinfo{booktitle}{\emph{In Proceedings of the International
  Workshop on Web Document Analysis (WDA 2001}}. \bibinfo{pages}{31--34}.
\newblock


\bibitem[\protect\citeauthoryear{Zhang, Zhang, and Balog}{Zhang
  et~al\mbox{.}}{2019}]%
        {table2vec}
\bibfield{author}{\bibinfo{person}{Li Zhang}, \bibinfo{person}{Shuo Zhang},
  {and} \bibinfo{person}{Krisztian Balog}.} \bibinfo{year}{2019}\natexlab{}.
\newblock \showarticletitle{Table2Vec: Neural Word and Entity Embeddings for
  Table Population and Retrieval}. In \bibinfo{booktitle}{\emph{Proceedings of
  the 42nd International ACM SIGIR Conference on Research and Development in
  Information Retrieval}}. \bibinfo{publisher}{Association for Computing
  Machinery}, \bibinfo{address}{New York, NY, USA},
  \bibinfo{pages}{1029–1032}.
\newblock
\showISBNx{9781450361729}


\bibitem[\protect\citeauthoryear{Zhang and Balog}{Zhang and Balog}{2017}]%
        {Zhang:2017:ESA}
\bibfield{author}{\bibinfo{person}{Shuo Zhang} {and} \bibinfo{person}{Krisztian
  Balog}.} \bibinfo{year}{2017}\natexlab{}.
\newblock \showarticletitle{EntiTables: Smart Assistance for Entity-Focused
  Tables}. In \bibinfo{booktitle}{\emph{Proceedings of the 40th International
  ACM SIGIR Conference on Research and Development in Information Retrieval}}
  (Shinjuku, Tokyo, Japan) \emph{(\bibinfo{series}{SIGIR '17})}.
  \bibinfo{publisher}{ACM}, \bibinfo{address}{New York, NY, USA},
  \bibinfo{pages}{255--264}.
\newblock
\showISBNx{978-1-4503-5022-8}
\urldef\tempurl%
\url{https://doi.org/10.1145/3077136.3080796}
\showDOI{\tempurl}


\bibitem[\protect\citeauthoryear{Zhang and Balog}{Zhang and Balog}{2018}]%
        {DBLP:journals/corr/abs-1802-06159}
\bibfield{author}{\bibinfo{person}{Shuo Zhang} {and} \bibinfo{person}{Krisztian
  Balog}.} \bibinfo{year}{2018}\natexlab{}.
\newblock \showarticletitle{Ad Hoc Table Retrieval using Semantic Similarity}.
  In \bibinfo{booktitle}{\emph{Proceedings of the 2018 World Wide Web
  Conference on World Wide Web, {WWW} 2018, Lyon, France, April 23-27, 2018}}.
  \bibinfo{publisher}{{ACM}}, \bibinfo{pages}{1553--1562}.
\newblock


\bibitem[\protect\citeauthoryear{Zhang and Balog}{Zhang and Balog}{2019a}]%
        {Zhang:2019:ADC}
\bibfield{author}{\bibinfo{person}{Shuo Zhang} {and} \bibinfo{person}{Krisztian
  Balog}.} \bibinfo{year}{2019}\natexlab{a}.
\newblock \showarticletitle{Auto-completion for Data Cells in Relational
  Tables}. In \bibinfo{booktitle}{\emph{Proceedings of the 28th ACM
  International Conference on Information and Knowledge Management}} (Beijing,
  China) \emph{(\bibinfo{series}{CIKM '19})}. \bibinfo{publisher}{ACM},
  \bibinfo{address}{New York, NY, USA}, \bibinfo{pages}{761--770}.
\newblock
\showISBNx{978-1-4503-6976-3}
\urldef\tempurl%
\url{https://doi.org/10.1145/3357384.3357932}
\showDOI{\tempurl}


\bibitem[\protect\citeauthoryear{Zhang and Balog}{Zhang and Balog}{2019b}]%
        {Zhang2019RecommendingRT}
\bibfield{author}{\bibinfo{person}{Shuo Zhang} {and} \bibinfo{person}{K.
  Balog}.} \bibinfo{year}{2019}\natexlab{b}.
\newblock \showarticletitle{Recommending Related Tables}.
\newblock \bibinfo{journal}{\emph{ArXiv}}  \bibinfo{volume}{abs/1907.03595}
  (\bibinfo{year}{2019}).
\newblock


\end{thebibliography}
%%
%% If your work has an appendix, this is the place to put it.
\appendix

\begin{figure}[t!]
    \centering
    \begin{subfigure}[t]{0.25\textwidth}
        \centering
        \includegraphics[height=1.5in]{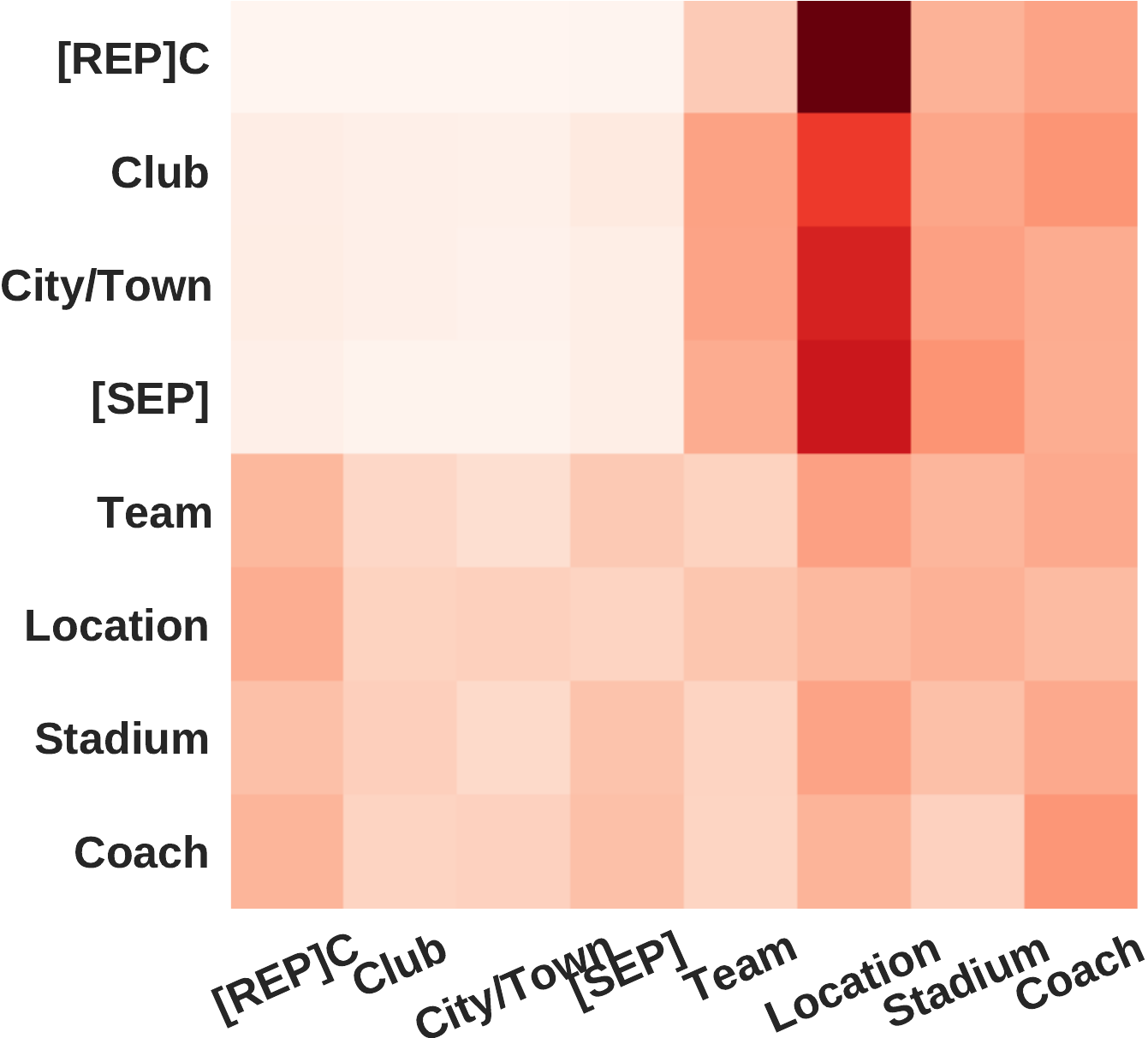}
        %\vspace*{-4mm} 
        \caption{1st attention head}
        
    \end{subfigure}%
    \begin{subfigure}[t]{0.25\textwidth}
        \centering
        \includegraphics[height=1.5in]{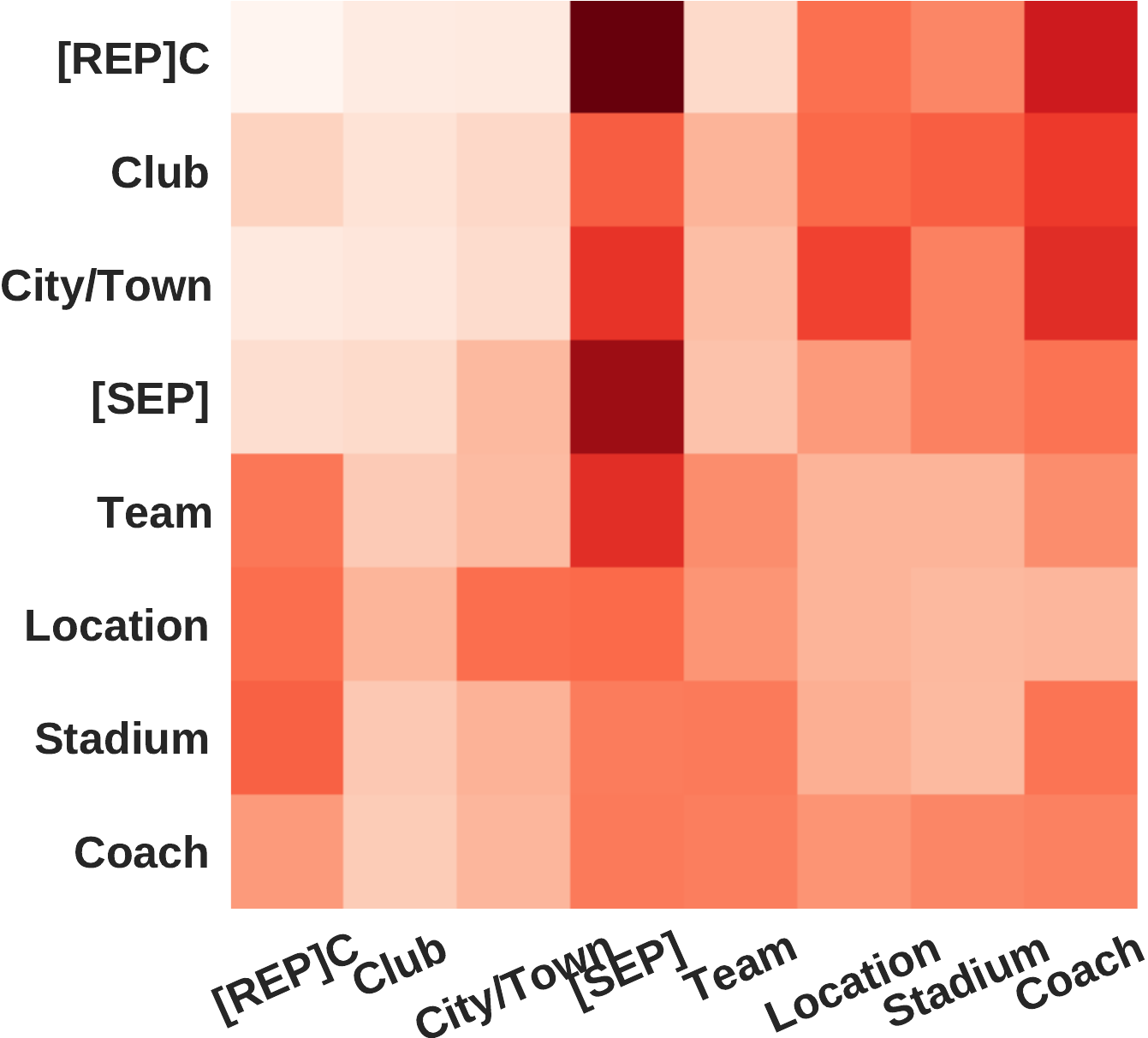}
        %\vspace*{-4mm} 
        \caption{2nd attention head}
    \end{subfigure}
    %\vspace*{-1mm}
    
    \begin{subfigure}[t]{0.25\textwidth}
        \centering
        \includegraphics[height=1.5in]{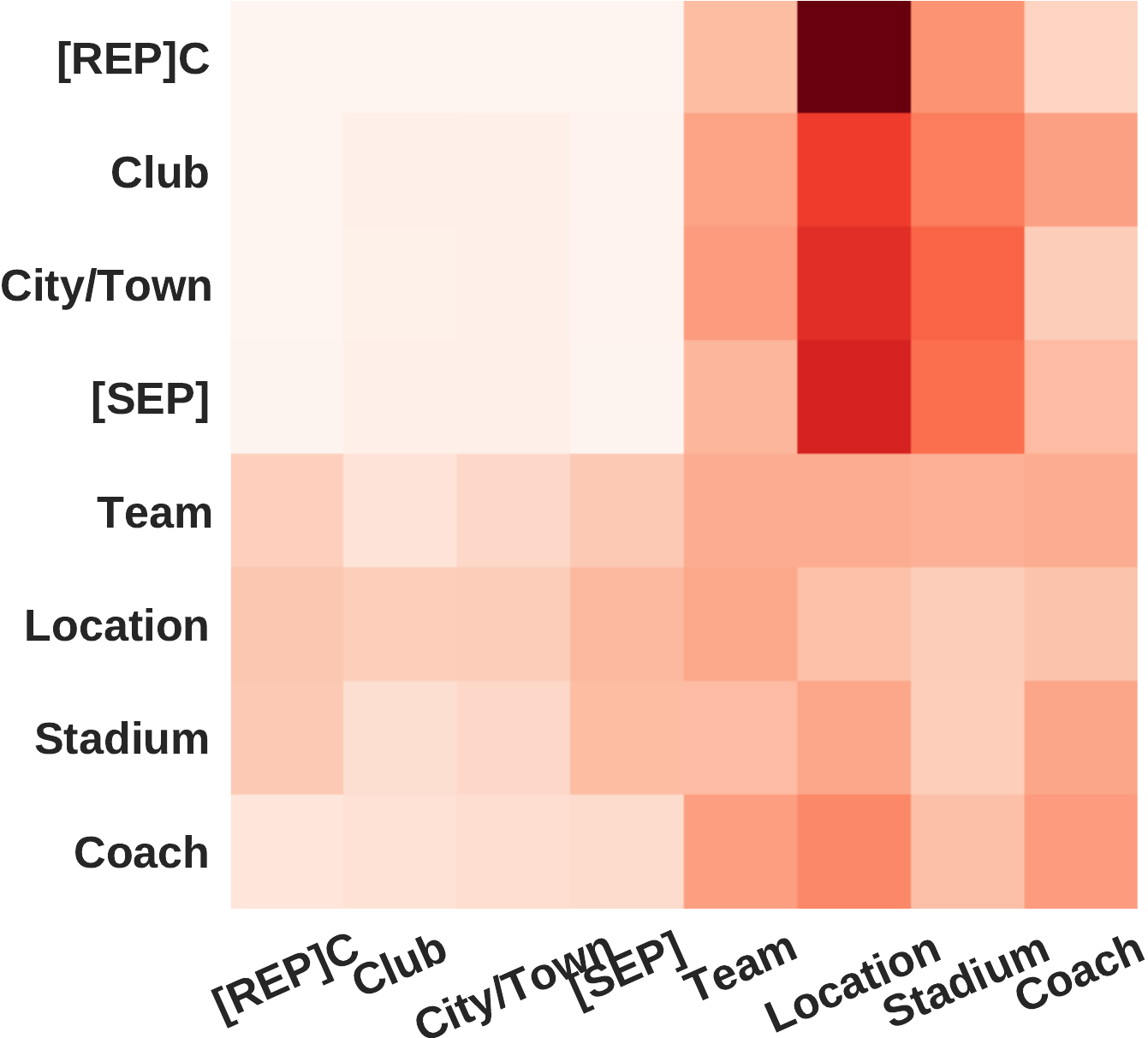}
        %\vspace*{-4mm} 
        \caption{3rd attention head}
        
    \end{subfigure}%
    \begin{subfigure}[t]{0.25\textwidth}
        \centering
        \includegraphics[height=1.5in]{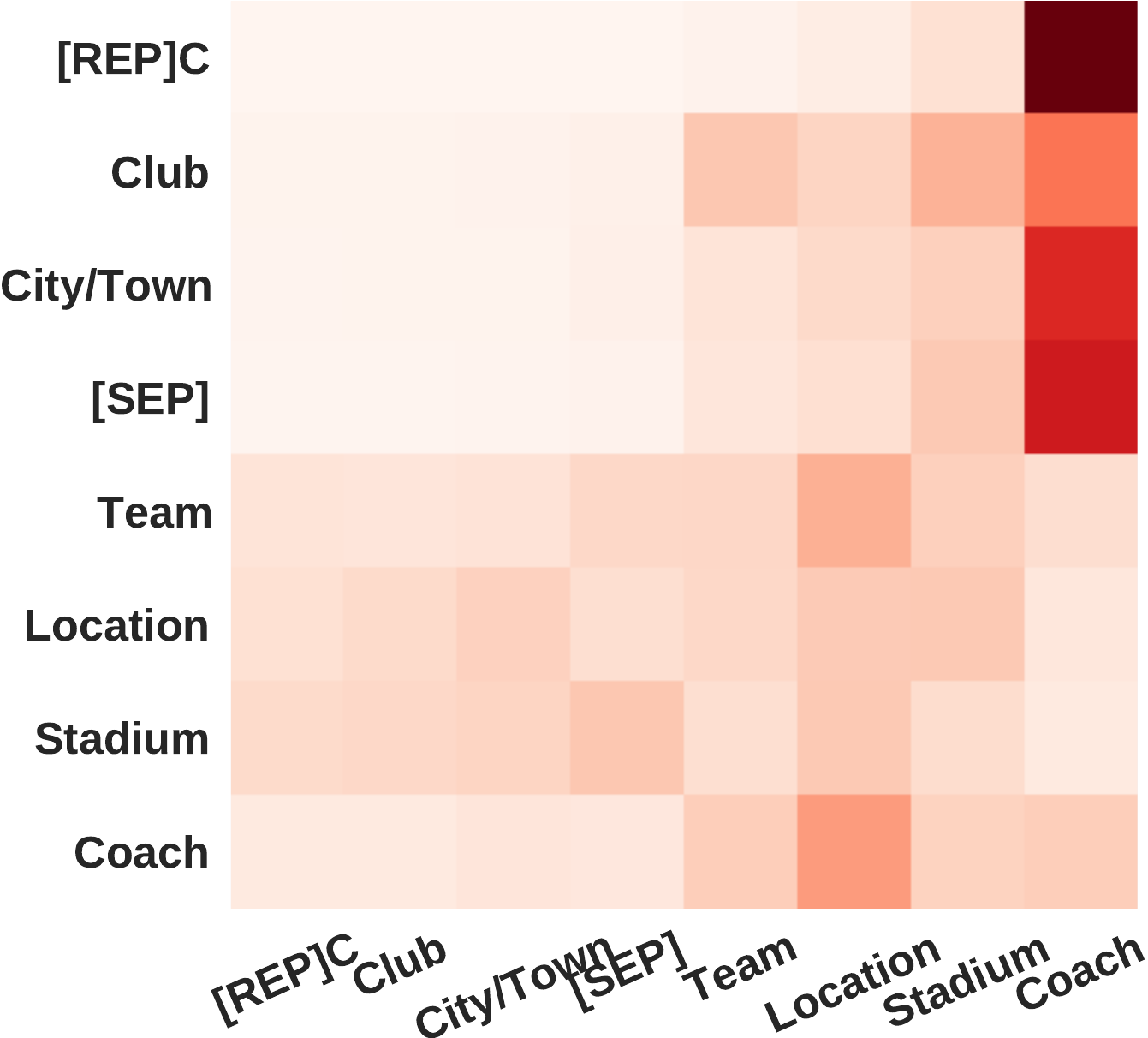}
        %\vspace*{-4mm} 
        \caption{4th attention head}
    \end{subfigure}
    \vspace*{-1mm}
    
    %\vspace*{-4mm}
    \caption{Comparison of attention heads between columns of a table pair.}
    \label{atention_heads}
    
\end{figure}

\section{miniBERT attention heads}

miniBERT is composed of one layer of Transformer blocks with four attention heads. To better understand how miniBERT works, we show the attention heads that correspond to a table similarity case. The first table is composed of the headers \textit{Club} and \textit{City/Town}, and the second table is composed of the headers \textit{Team}, \textit{Location}, \textit{Stadium}, and \textit{Coach}. Figure \ref{atention_heads} illustrates the four attention heads of miniBERT. Figure \ref{atention_heads}(a) indicates that the 1st attention head focuses on the header \textit{Location} from the second table which attends mainly to the header \textit{City/Town} from the first table and contributes significantly to the embedding $\boldsymbol{[REP]_c}$. The second attention head, illustrated in Figure \ref{atention_heads}(b), is more general as it indicates multiple cross matching signals between columns of both tables. The third attention head in Figure \ref{atention_heads}(c) is similar to the 1st attention head with more focus on the header \textit{Stadium} from the second table. This can be explained by the co-occurrence of the header \textit{Stadium} with headers \textit{Club} and \textit{City/Town}. We also observe similar patterns in the 4th attention head that focuses mainly on the header \textit{Coach}. The analysis of the attention heads shows the advantage of using the Transformers blocks to capture the many-to-many relationships between columns of tables by aggregating information both within and across table columns.

\end{document}